\documentclass[twocolumn,english,amsmath,amssymb,nofootinbib,superscriptaddress,showpacs,showkeys]{revtex4}
\usepackage{ae,aecompl}
\usepackage[T1]{fontenc}
\usepackage[latin1]{inputenc}
\usepackage{color}
\definecolor{shadecolor}{rgb}{1, 0, 0}
\usepackage{babel}
\usepackage{framed}
\usepackage{bm}
\usepackage{amsmath}
\usepackage{graphicx}
\usepackage{esint}
\usepackage[unicode=true,
 bookmarks=false,
 breaklinks=false,pdfborder={0 0 1},backref=false,colorlinks=false]
 {hyperref}

\makeatletter
\@ifundefined{textcolor}{}
{%
 \definecolor{BLACK}{gray}{0}
 \definecolor{WHITE}{gray}{1}
 \definecolor{RED}{rgb}{1,0,0}
 \definecolor{GREEN}{rgb}{0,1,0}
 \definecolor{BLUE}{rgb}{0,0,1}
 \definecolor{CYAN}{cmyk}{1,0,0,0}
 \definecolor{MAGENTA}{cmyk}{0,1,0,0}
 \definecolor{YELLOW}{cmyk}{0,0,1,0}
}

\usepackage{babel}
\usepackage{babel}
\usepackage{babel}
\usepackage{babel}
\usepackage{babel}

\usepackage{aecompl}

\def\Phid{P_{{\rm hid}}}
\def\CovU{\nabla_{U}}
\def\LieU{\mathcal{L}_{U}}
\def\OmOrb{\omega}
\def\OmHel{\Omega}

\def\be{\begin{equation}}
\def\ee{\end{equation}}
\def\best{\begin{equation*}}
\def\eest{\end{equation*}}
\def\beqn{\begin{eqnarray}}
\def\eeqn{\end{eqnarray}}

\makeatother

\begin{document}

\title{Spinning particles in general relativity: \\
Momentum-velocity relation for the Mathisson-Pirani spin condition}

\author{L. Filipe O. Costa}

\email{lfpocosta@math.ist.utl.pt}

\selectlanguage{english}%

\affiliation{Center for Mathematical Analysis, Geometry and Dynamical Systems,
Instituto Superior Técnico, Universidade de Lisboa, 1049-001 Lisboa,
Portugal}

\author{Georgios Lukes-Gerakopoulos}

\email{gglukes@gmail.com}

\selectlanguage{english}%

\affiliation{Astronomical Institute of the Academy of Sciences of the Czech Republic,
Bo\v{c}n\'{i} II 1401/1a, CZ-141 31 Prague, Czech Republic }

\author{Old\v{r}ich Semerák}

\email{semerak@mbox.troja.mff.cuni.cz}

\selectlanguage{english}%

\affiliation{Institute of Theoretical Physics, Faculty of Mathematics and Physics,
Charles University, CZ-180 00 Prague, Czech Republic}

\date{\today}
\begin{abstract}
The Mathisson-Papapetrou-Dixon (MPD) equations, providing the ``pole-dipole''
description of spinning test particles in general relativity, have
to be supplemented by a condition specifying the worldline that will
represent the history of the studied body. It has long been thought
that the Mathisson-Pirani (MP) spin condition---unlike other major
choices made in the literature---does {\em not} yield an explicit
momentum-velocity relation. We derive here the desired (and very simple)
relation and show that it is in fact {\em equivalent} to the MP
condition. We clarify the apparent paradox between the existence of
such a definite relation and the known fact that the MP condition
is degenerate (does not specify a unique worldline), thus shedding
light on some conflicting statements made in the literature. We then
show how, for a given body, this spin condition yields infinitely
many possible representative worldlines, and derive a detailed method
how to switch between them in a curved spacetime. The MP condition
is a convenient choice in situations when it is easy to recognize
its ``nonhelical'' solution, as exemplified here by bodies in circular
orbits and in radial fall in the Schwarzschild spacetime.
\end{abstract}
\maketitle
\tableofcontents{}

\section{Introduction}

The problem of motion of a ``small body'' in general relativity
has been widely studied in the ``pole-dipole'' test-particle approximation
when the body is not itself contributing to the gravitational field
and when it is only characterized by mass and spin (proper angular
momentum), with all the higher multipoles neglected. If the particle
interacts solely gravitationally, the only force it is subjected to
comes from the spin-curvature interaction and the pole-dipole problem
is described by the Mathisson--Papapetrou--Dixon (MPD) equations 
\begin{align}
\frac{{\rm D}P^{\mu}}{{\rm d}\tau} & =-\frac{1}{2}\,{R^{\mu}}_{\nu\kappa\lambda}U^{\nu}S^{\kappa\lambda}\ \equiv\ F^{\mu},\label{Papa-p}\\
\frac{{\rm D}S^{\alpha\beta}}{{\rm d}\tau} & =2P^{[\alpha}U^{\beta]}\equiv P^{\alpha}U^{\beta}-U^{\alpha}P^{\beta}\quad,\label{eq:Spinevol}
\end{align}
where $P^{\mu}$ and $S^{\mu\nu}$ denote, respectively, the body's
4-momentum and spin tensor (spin bivector), ${\displaystyle U^{\mu}\equiv{\rm d}z^{\mu}/{\rm d}\tau}$
the 4-velocity of the body's representative worldline $z^{\mu}(\tau)$,
and 
\[
{\displaystyle \frac{{\rm D}}{{\rm d}\tau}\equiv\CovU\equiv~_{;\mu}U^{\mu}}
\]
denotes the covariant derivative along $U^{\mu}$. Both $P^{\mu}$
and $U^{\mu}$ are assumed to be timelike, with $U^{\mu}$ normalized
to $U_{\mu}U^{\mu}\!=\!-1$, which implies that $\tau$ is the proper
time. Contractions of $P^{\mu}$ and $U^{\mu}$ provide the masses
$\mathcal{M}$ and $m$, 
\[
-P_{\mu}P^{\mu}\equiv\mathcal{M}^{2}>0\,;\quad-P_{\mu}U^{\mu}\equiv m>0\,,
\]
respectively, the mass as measured in the zero 3-momentum and in the
zero 3-velocity frames. The timelike character of both $P^{\mu}$
and $U^{\mu}$ is however not guaranteed automatically by the MPD
equations, with possible breakdown of this requirement indicating
ultimate limits of the pole-dipole description. The spin bivector
is assumed to be spacelike, so 
\[
\frac{1}{2}\,S_{\mu\nu}S^{\mu\nu}\equiv S^{2}>0\,.
\]

Since the MPD set (\ref{Papa-p})-(\ref{eq:Spinevol}) possesses 13
unknowns\footnote{Four independent components of \textcolor{black}{$P^{\alpha}$, 3
independent components of $U^{\alpha}$, and 6 independent components
of $S^{\alpha\beta}$.}} for only 10 equations, in order to be closed, it has to be supplemented
by 3 auxiliary constraints. These are provided by the so-called spin
supplementary condition (SSC), standardly written as 
\[
S^{\mu\nu}V_{\nu}\!=\!0\ ,
\]
where, in case of a particle with nonzero rest mass, $V^{\mu}$ is
some (freely chosen) timelike vector field {[}defined at least along
$z^{\mu}(\tau)${]} which is supposed to normalize as $V_{\mu}V^{\mu}\!=\!-1$.
This condition is a choice of a representative worldline $z^{\mu}(\tau)$;
more precisely, it demands $z^{\mu}(\tau)$ to be, at each instant,
the body's center of mass (or ``centroid'') as measured by an observer
with instantaneous 4-velocity $V^{\mu}$. Four choices of $V^{\mu}$
have proven particularly convenient: 
\begin{itemize}
\item $V^{\mu}\!\equiv\!U^{\mu}$ (Mathisson-Pirani (MP) condition \cite{Mathisson37,Pirani56},
originally due to Frenkel \cite{Frenkel}), which states that the
reference worldline $z^{\mu}(\tau)$ is the centroid as measured in
its own rest frame (the zero 3-velocity frame); 
\item $V^{\mu}\!\equiv\!P^{\mu}/\mathcal{M}$ (Tulczyjew-Dixon (TD) condition,
\cite{TulczyjewII,Dixon1970I}), which states that $z^{\mu}(\tau)$
is the centroid as measured in the zero 3-momentum frame; 
\item $V^{\mu}\propto u_{{\rm lab}}^{\mu}+P^{\mu}/\mathcal{M}$ (Newton-Wigner
(NW) condition, \cite{Newton-Wigner,Pryce}), where $u_{{\rm lab}}^{\mu}\propto\partial_{t}^{\mu}$
is the 4-velocity of the congruence of ``laboratory'' observers,
at rest in the given coordinate system (typically somehow privileged
by symmetries of the host spacetime); 
\item $P^{\mu}\!=\!mU^{\mu}$ ($P^{\mu}\!\parallel\!U^{\mu}$ condition,
known also as Ohashi-Kyrian-Semerák (OKS) condition \cite{Ohashi-03,Semerak II}),
which demands $V^{\mu}$ to be such that ${\rm D}V^{\mu}/{\rm d}\tau$
belongs to the eigenplane of $S^{\mu\nu}$ \cite{SemerakS-PRDI},
for instance when $V^{\mu}$ parallel transports along $z^{\mu}(\tau)$,
${\rm D}V^{\mu}/{\rm d}\tau=0$. 
\end{itemize}
A fifth, less popular choice, is $V^{\mu}\equiv u_{{\rm lab}}^{\mu}$
(Corinaldesi-Papapetrou (CP) condition, \cite{Corinaldesi51}), which
states that $z^{\mu}(\tau)$ is the centroid as measured in the ``laboratory''
frame \cite{CostaNatario2014}.

The TD choice has been used most frequently, mainly because it leads
to an explicit expression of the tangent $U^{\mu}$ in terms of $P^{\mu}$,
$S^{\mu\nu}$, and $z^{\mu}$, the so-called momentum-velocity relation
\cite{Kunzle-72}, 
\begin{equation}
U^{\mu}=\frac{m}{\mathcal{M}^{2}}\left(P^{\mu}+\frac{2S^{\mu\nu}R_{\nu\iota\kappa\lambda}P^{\iota}S^{\kappa\lambda}}{4\mathcal{M}^{2}+R_{\alpha\beta\gamma\delta}S^{\alpha\beta}S^{\gamma\delta}}\right)\quad.\label{eq:P_U_Dixon}
\end{equation}
Such relation is important, mainly in numerical treatment, where the
integration of the MPD system is done recurrently using the instantaneous
tangent $U^{\mu}$ (see \cite{Semerak I} for details). The $P^{\mu}\!=\!mU^{\mu}$
option in itself represents the momentum-velocity relation and it
turned out to simplify the spinning-particle problem considerably
\cite{SemerakS-PRDI}. For the CP and NW conditions a momentum-velocity
relation is obtainable, but complicated \cite{CostaNatario2014},
and no explicit expression has yet been put forth. Finally, the MP
SSC has also been used many times, but it has been thought that it
does not lead to an explicit momentum-velocity relation (it has only
been shown to provide such an expression for the four-acceleration
${\displaystyle {\rm D}U^{\mu}/{\rm d}\tau}$, \cite{Semerak II}).
For recent discussions of the subject, see e.g. \cite{CostaNatario2014,SemerakS-PRDI}.

\paragraph*{Units and notation:}

Geometric units are used throughout the article, ${G=c=1}$. Greek
letters denote the indices corresponding to spacetime, while Latin
letters denote indices corresponding only to space. We use the Riemann
tensor convention ${{R^{\alpha}}_{\beta\gamma\delta}=\partial_{\gamma}\Gamma_{\beta\delta}^{\alpha}-\partial_{\delta}\Gamma_{\beta\gamma}^{\alpha}+...}$,
with metric signature $(-,+,+,+)$. We use abstract index notation
for tensors $T^{\alpha\beta\gamma...}$ and 4-vectors $V^{\alpha}$;
arrow notation $\vec{V}$ denotes space components of a vector in
a given frame. The Levi-Civita tensor is $\epsilon_{\mu\nu\rho\sigma}=\sqrt{-g}\tilde{\epsilon}_{\mu\nu\rho\sigma}$,
with the Levi-Civita symbol $\tilde{\epsilon}_{0123}=1$.

\section{Equations of motion under a spin supplementary condition (SSC)}

First, just for self-completeness, let us repeat several simple formulas
from \cite{SemerakS-PRDI,CostaNatarioZilhao}. Writing, for a {\em
general} vector $V^{\mu}$, the spin bivector in terms of the corresponding
spin vector $S^{\mu}\equiv-\epsilon_{\ \nu\alpha\beta}^{\mu}V^{\nu}S^{\alpha\beta}/2$,
\begin{equation}
S_{\alpha\beta}=\epsilon_{\alpha\beta\gamma\delta}V^{\gamma}S^{\delta}\,,\label{eq:Stensor-Svector}
\end{equation}
its evolution along $U^{\mu}$ is just 
\begin{equation}
\frac{{\rm D}S_{\alpha\beta}}{{\rm d}\tau}=\epsilon_{\alpha\beta\gamma\delta}\frac{{\rm D}V^{\gamma}}{{\rm d}\tau}S^{\delta}+\epsilon_{\alpha\beta\gamma\delta}V^{\gamma}\frac{{\rm D}S^{\delta}}{{\rm d}\tau}\quad.\label{dotS-dots}
\end{equation}
In order to ``extract'' the evolution of $S^{\mu}$, one substitutes
Eq.~\eqref{eq:Spinevol} in Eq.~\eqref{dotS-dots} and multiplies
this with $\epsilon^{\mu\nu\alpha\beta}V_{\nu}$, i.e. 
\begin{equation}
(\delta_{\nu}^{\mu}+V^{\mu}V_{\nu})\,\frac{{\rm D}S^{\nu}}{{\rm d}\tau}=\epsilon^{\mu\nu\alpha\beta}V_{\nu}U_{\alpha}P_{\beta}\quad,\label{dots,project}
\end{equation}
and hence 
\begin{equation}
\frac{{\rm D}S^{\mu}}{{\rm d}\tau}=V^{\mu}\frac{{\rm D}V_{\nu}}{{\rm d}\tau}S^{\nu}+\epsilon^{\mu\nu\alpha\beta}V_{\nu}U_{\alpha}P_{\beta}\,\quad.\label{dots}
\end{equation}
This yields 
\begin{align}
S\,\frac{{\rm d}S}{{\rm d}\tau}=\frac{1}{2}\frac{{\rm d}S^{2}}{{\rm d}\tau} & =S_{\mu}\frac{{\rm D}S^{\mu}}{{\rm d}\tau}=\epsilon^{\mu\nu\alpha\beta}V_{\mu}S_{\nu}P_{\alpha}U_{\beta}\label{sds}
\end{align}
for evolution of the spin magnitude $S=\sqrt{S_{\mu}S^{\mu}}=\sqrt{S_{\alpha\beta}S^{\alpha\beta}/2}$.

In order to express the evolution of $V^{\mu}$ instead, one multiplies
the relation~\eqref{dotS-dots} by $\epsilon^{\mu\nu\alpha\beta}S_{\nu}$
and uses Eq.~\eqref{sds}, to arrive at 
\begin{equation}
(S^{2}\delta_{\nu}^{\mu}-S^{\mu}S_{\nu})\,\frac{{\rm D}V^{\nu}}{{\rm d}\tau}=(\delta_{\iota}^{\mu}+V^{\mu}V_{\iota})\,\epsilon^{\iota\nu\alpha\beta}S_{\nu}U_{\alpha}P_{\beta}\quad,\label{dotV,project}
\end{equation}
and hence 
\begin{equation}
{\displaystyle S\,\frac{{\rm D}(S~V^{\mu})}{{\rm d}\tau}=-S^{\mu}\frac{{\rm D}S_{\nu}}{{\rm d}\tau}V^{\nu}+\epsilon^{\mu\nu\alpha\beta}S_{\nu}U_{\alpha}P_{\beta}\,\quad.}\label{dotV}
\end{equation}
Substituting Eq. (\ref{eq:Stensor-Svector}) into (\ref{Papa-p}),
we can also express the force in terms of the spin vector $S^{\mu}$,
\begin{equation}
F^{\alpha}\equiv\frac{{\rm D}P^{\alpha}}{{\rm d}\tau}=\star R^{\sigma\tau\alpha\mu}S_{\sigma}V_{\tau}U_{\mu}\ ,\label{eq:ForceSGen}
\end{equation}
where $\star R_{\alpha\beta\gamma\delta}\equiv\epsilon_{\alpha\beta}^{\ \ \mu\nu}R_{\mu\nu\gamma\delta}/2$.

Let us stress again that, up to now, everything has been valid for
a {\em generic} timelike vector $V^{\mu}$. For this generic vector
one can obtain a general $P-U$ relation by contracting the spin evolution
equation~\eqref{eq:Spinevol} with $V_{\beta}$, and noticing that,
by virtue of $S^{\alpha\beta}V_{\beta}=0$, $V_{\beta}{\rm D}S^{\alpha\beta}/{\rm d}\tau=-S^{\alpha\beta}{\rm D}V_{\beta}/{\rm d}\tau$,
leading to \cite{Buonanno2009,Steinhoff2011,Helical,SteinhoffPuetzfeld2012}
\begin{equation}
P^{\alpha}=\frac{1}{\gamma(V,U)}\left(\mu~U^{\alpha}+S^{\alpha\beta}\frac{{\rm D}V_{\beta}}{{\rm d}\tau}\right)\ \quad,\label{eq:Momentum}
\end{equation}
where $\mu\equiv-P^{\alpha}V_{\alpha}$ is the mass as measured by
an observer of 4-velocity $V^{\alpha}$, and $\gamma(V,U)\equiv-U^{\alpha}V_{\alpha}$
is the Lorentz factor between $U^{\alpha}$ and $V^{\alpha}$.

\subsection{MPD system under the MP SSC}

Consider now the MP SSC, i.e., let $V^{\mu}\!\equiv\!U^{\mu}$. The
force equation (\ref{eq:ForceSGen}) becomes \cite{CostaNatarioZilhao}\footnote{There is a sign difference compared to the expression in \cite{CostaNatarioZilhao},
due to the different sign convention for the Levi-Civita tensor. } 
\begin{equation}
F^{\alpha}\equiv\frac{{\rm D}P^{\alpha}}{{\rm d}\tau}=\mathbb{H}^{\beta\alpha}S_{\beta}\quad,\label{eq:TidaltensorF}
\end{equation}
where $\mathbb{H}_{\ \beta}^{\alpha}\equiv\star R_{\ \mu\beta\nu}^{\alpha}U^{\mu}U^{\nu}=\epsilon_{\ \mu\sigma\tau}^{\alpha}R_{\ \ \beta\nu}^{\sigma\tau}U^{\mu}U^{\nu}/2$
is the ``gravitomagnetic tidal tensor'' (or ``magnetic part of
the Riemann tensor'') as measured by an observer of 4-velocity $U^{\alpha}$.
The spin evolution equation becomes the Fermi-Walker transport law
(e.g. \cite{Gravitation}), 
\begin{align}
\frac{{\rm D}S^{\mu}}{{\rm d}\tau}=U^{\mu}\frac{{\rm D}U^{\nu}}{{\rm d}\tau}S_{\nu} & \ .\label{eq:MPFermi}
\end{align}
These expressions are a unique feature of the MP SSC. Equation (\ref{eq:MPFermi})
tells us that $S^{\mu}$ has fixed components in the locally nonrotating
frame comoving with the centroid. A locally nonrotating frame is mathematically
defined as Fermi-Walker transported frame, and is physically realized
as a frame where the Coriolis forces vanish. This means that $S^{\alpha}$
follows the ``compass of inertia'' \cite{MassaZordan}, which is
the most natural spin behavior (in the absence of torques), since
gyroscopes are well known for opposing to changes in direction of
their rotation axis. (The spin vectors of other spin conditions, by
contrast, are not fixed, in general, relative to the comoving nonrotating
frame).

\subsection{The momentum-acceleration relation for the MP SSC}

For $V^{\mu}=U^{\mu}$, Eq. \eqref{sds} implies that $S$ is a constant,
thus Eq.~\eqref{dotV} is rewritten as 
\begin{align}
S^{2}\frac{{\rm D}U^{\mu}}{{\rm d}\tau} & =-S^{\mu}\frac{{\rm D}S_{\nu}}{{\rm d}\tau}U^{\nu}+\epsilon^{\mu\nu\alpha\beta}S_{\nu}U_{\alpha}P_{\beta}\nonumber \\
 & =-S^{\mu}\frac{{\rm D}S_{\nu}}{{\rm d}\tau}U^{\nu}-S^{\mu\beta}P_{\beta}\quad,\label{dotu}
\end{align}
since now 
\begin{align}
S^{\mu\beta}=\epsilon^{\mu\beta\alpha\nu}U_{\alpha}S_{\nu};\qquad S^{\alpha}=-\frac{1}{2}\epsilon^{\alpha\beta\gamma\delta}U_{\beta}S_{\gamma\delta}\ \ .\label{eq:Sten2Svec}
\end{align}
Note that the first term on the right-hand side of Eq.~\eqref{dotu}
can also be rewritten as $-\frac{1}{m}\,S^{\mu}P^{\nu}{\rm D}S_{\nu}/d\tau$
thanks to the generally valid relation 
\[
\gamma(V,U)\,P^{\nu}\frac{{\rm D}S_{\nu}}{{\rm d}\tau}\!=\!\mu\,U^{\nu}\frac{{\rm D}S_{\nu}}{{\rm d}\tau}
\]
(which specifically for $V^{\mu}\!\equiv\!U^{\mu}$ means ${\displaystyle P^{\nu}{\rm D}S_{\nu}/{\rm d}\tau\!=\!mU^{\nu}{\rm D}S_{\nu}/{\rm d}\tau}$).
One, thus, obtains the momentum-acceleration relation reached in \cite{Semerak II},
\begin{equation}
a^{\alpha}\equiv\frac{{\rm D}U^{\alpha}}{{\rm d}\tau}=\frac{1}{S^{2}}\left(\frac{1}{m}F^{\mu}S_{\mu}S^{\alpha}-P_{\gamma}S^{\alpha\gamma}\right)\ \quad,\label{eq:ExplicitaFMP}
\end{equation}
where $a^{\alpha}$ is the acceleration.

\subsection{The momentum-velocity relation for the MP SSC}

For $V^{\mu}=U^{\mu}$, Eq. (\ref{eq:Momentum}) yields 
\begin{equation}
P^{\mu}=m\,U^{\mu}+S^{\mu\nu}a_{\nu}\,\quad,\label{p,u,dSu}
\end{equation}
where $a^{\mu}={\rm D}U^{\mu}/{\rm d}\tau$. The desired momentum-velocity
relation\footnote{Equation (\ref{eq:momentumVelMP}) has recently been obtained, for
the \emph{special case} of flat spacetime, in \cite{CostaNatario2014}.
Herein we show it to be valid in general.} follows simply by substituting in Eq.~\eqref{p,u,dSu} the acceleration
$a_{\nu}$ from Eq.~\eqref{dotu}, 
\begin{equation}
m\,U^{\mu}=P^{\mu}+\frac{1}{S^{2}}\,S^{\mu\nu}S_{\nu\beta}\,P^{\beta}\,\quad.\label{eq:momentumVelMP}
\end{equation}
(One only employs the fact that $S^{\mu\nu}S_{\nu}\!=\!0$ by definition.)
The relation (\ref{eq:momentumVelMP}) contains two scalars, $m$
and $S$, which are constant in case of the MP SSC. Therefore, they
are fixed by the initial conditions. Contracting Eq. (\ref{eq:momentumVelMP})
with $P^{\alpha}$, we get 
\begin{equation}
m^{2}=\mathcal{M}^{2}-\frac{1}{S^{2}}S^{\alpha\mu}S_{\mu\beta}P^{\beta}P_{\alpha}\ ;\label{eq:MassMP}
\end{equation}
substituting back into Eq.~\eqref{eq:momentumVelMP} leads to an
explicit equation for $U^{\alpha}$ in terms of $P^{\mu}$ and $S^{\mu\nu}$
only, $U^{\alpha}=U^{\alpha}(P^{\mu},S^{\mu\nu})$.

The existence of the relation~\eqref{eq:momentumVelMP} might seem
strange, mainly due to the long history of assertions that no such
relation is available for the MP condition. Such assertions were actually
followed by a debate on the freedom in choosing initial conditions
and on the subsequent option for ``helical'' motion. These issues
shall be discussed in detail in Sec.~\ref{sec:The-duality-between}.

\subsubsection{Simple checks of the relation}

As a first check, we note that, since $S^{\mu\nu}S_{\nu}^{\ \beta}=S^{\mu}S^{\beta}-h^{\mu\beta}S^{2}$,
where 
\begin{equation}
h_{\ \beta}^{\mu}\equiv\delta_{\ \beta}^{\mu}+U^{\mu}U_{\beta}\label{eq:SpaceProj}
\end{equation}
is the space projector orthogonal to $U^{\mu}$, substituting into
(\ref{eq:momentumVelMP}) yields the trivial relation $mU^{\mu}=P^{\mu}-h_{\ \beta}^{\mu}P^{\beta}$
($\Leftrightarrow mU^{\mu}=mU^{\mu}$), stating that $mU^{\mu}$ is
the component of $P^{\mu}$ parallel to $U^{\mu}$. That (\ref{eq:momentumVelMP})
verifies the 4-velocity normalization also follows trivially from
this relation.

Let us imagine now that the relation~\eqref{eq:momentumVelMP} is
considered in a generic case, {\em without specifying any spin condition}.
It is (in any case) useful to express 
\begin{align}
S^{\mu\nu}S_{\nu\beta} & =\epsilon^{\mu\nu\kappa\lambda}V_{\kappa}S_{\lambda}\,\epsilon_{\nu\beta\rho\sigma}V^{\rho}S^{\sigma}=\nonumber \\
 & =S^{2}\left(-\delta_{\beta}^{\mu}-V^{\mu}V_{\beta}+S^{-2}S^{\mu}S_{\beta}\right)\quad,\label{SS}
\end{align}
and thus to rewrite Eq.~\eqref{eq:momentumVelMP} as 
\begin{equation}
m\,U^{\mu}=\left(-V^{\mu}V_{\beta}+S^{-2}S^{\mu}S_{\beta}\right)P^{\beta}\,\quad,\label{up-relation,Vs}
\end{equation}
which reveals that geometrically it means projection of $P^{\mu}$
on the eigenplane of $S^{\mu\nu}$ (or, equivalently, on the blade
of its dual bivector). Note that since $S_{\beta}U^{\beta}\!=\!0$
$\Leftrightarrow$ $S_{\beta}P^{\beta}\!=\!0$ (this is generally
valid, see Eq. (\ref{eq:Momentum})) and the former {\em is} true
if the MP condition holds, the relation reduces to trivial $mU^{\mu}\!=\!mU^{\mu}$
in that case.

An important property is evident now: if multiplied by $S_{\alpha\mu}$,
relation~\eqref{up-relation,Vs} gives 
\[
m\,S_{\alpha\mu}U^{\mu}=S_{\alpha\mu}\left(-V^{\mu}V_{\beta}+S^{-2}S^{\mu}S_{\beta}\right)P^{\beta}=0
\]
immediately, because $S_{\alpha\mu}V^{\mu}\!=\!0$ as well as $S_{\alpha\mu}S^{\mu}\!=\!0$
by definition. Therefore, relation~\eqref{eq:momentumVelMP} implies
the MP SSC, and so (since the MP SSC likewise implies \eqref{eq:momentumVelMP})
it is {\em equivalent} to the latter.

\subsubsection{``Hidden momentum''}

The component of $P^{\mu}$ orthogonal to $U^{\mu}$, $h_{\ \nu}^{\mu}P^{\nu}\equiv P_{{\rm hid}}^{\mu}$,
has been dubbed in some literature ``hidden momentum.'' The reason
for the denomination is seen taking the perspective of an observer
comoving with the centroid (the zero 3-velocity frame). In such frame
the spatial momentum is precisely $h_{\ \nu}^{\mu}P^{\nu}$, and is
in general nonzero. However, by definition, the body is ``at rest''
in this frame (since this is the rest frame of the center of mass,
or centroid, chosen to represent it); hence such momentum must be
hidden somehow. It may be cast as analogous (albeit with a very different
nature \cite{Wald et al 2010,CostaNatarioZilhao}) to the hidden momentum
first found in electromagnetic systems \cite{Shockley} (namely in
magnetic dipoles subjected to electric fields \cite{Shockley,GriffithsAmJPhys,Vaidman,Wald et al 2010,CostaNatarioZilhao}).
The concept proved useful in simplifying the interpretation of some
exotic motions of the centroid in Refs. \cite{Wald et al 2010,Helical,CostaNatario2014}
(amongst them the Mathisson helical motions \cite{Helical}, discussed
below). It reads, for the MP condition, 
\begin{align}
\Phid^{\mu} & \equiv\ h_{\ \alpha}^{\mu}\,P^{\alpha}=P^{\mu}-mU^{\mu}\label{eq:HidddenMom0}\\
 & =\ S^{\mu\nu}a_{\nu}\ =\ -\epsilon_{\ \beta\gamma\delta}^{\mu}S^{\beta}a^{\gamma}U^{\delta}\,\quad,\label{eq:hiddenMom}
\end{align}
the last two equalities holding for the MP SSC, where we used Eqs.
(\ref{p,u,dSu}), (\ref{eq:Sten2Svec}). Relation~\eqref{eq:momentumVelMP}
yields an alternative expression for the hidden momentum, in terms
of $P^{\mu}$ and $S^{\mu\nu}$: 
\begin{align}
\Phid^{\mu}=-\frac{1}{S^{2}}\,S^{\mu\nu}S_{\nu\beta}\,P^{\beta}\,\quad.\label{eq:hiddenMom2}
\end{align}

\section{The duality between the degeneracy of centroid and the determinacy
of the equations\label{sec:The-duality-between}}

Equations (\ref{eq:momentumVelMP})-(\ref{eq:MassMP}) yield a momentum-velocity
relation of the form $U^{\alpha}=U^{\alpha}(P^{\mu},S^{\mu\nu})$.
This means that the equations of motion can be written as the \emph{explicit}
functions 
\begin{align}
\frac{{\rm d}z^{\alpha}}{{\rm d}\tau} & =U^{\alpha}(z^{\mu},P^{\mu},S^{\mu\nu});\nonumber \\
\frac{{\rm d}P^{\alpha}}{{\rm d}\tau} & =f^{\alpha}(z^{\mu},P^{\mu},S^{\mu\nu});\label{eq:EqsMotion}\\
{\displaystyle \frac{{\rm d}S^{\alpha\beta}}{{\rm d}\tau}} & =\psi^{\alpha\beta}(z^{\mu},P^{\mu},S^{\mu\nu})\nonumber 
\end{align}
which, given the initial values $\{z^{\alpha},P^{\alpha},S^{\alpha\beta}\}|_{{\rm in}}$,
form a determinate system. That is, the solution is unique given this
type of initial data and hence, from this point of view, the MP SSC
works like the other SSC's in the literature. This fact might be surprising
at first, since, contrary to other SSCs like the TD, the MP SSC is
known for not specifying a unique worldline through the body \cite{MollerAIP,TulczyjewII,Dixon1970I,Semerak II,Wald et al 2010,Steinhoff2011,CostaNatario2014},
being infinitely degenerated. This led to (apparent) contradictions
in the literature, between authors noticing that it does not uniquely
specify a worldline, and those arguing \cite{Obukhov,ObukhovPuetzfeld}
that it does, given certain initial conditions (see in particular
the comments made in \cite{Obukhov}).

\begin{figure}[h!!!]
\includegraphics[width=1\columnwidth]{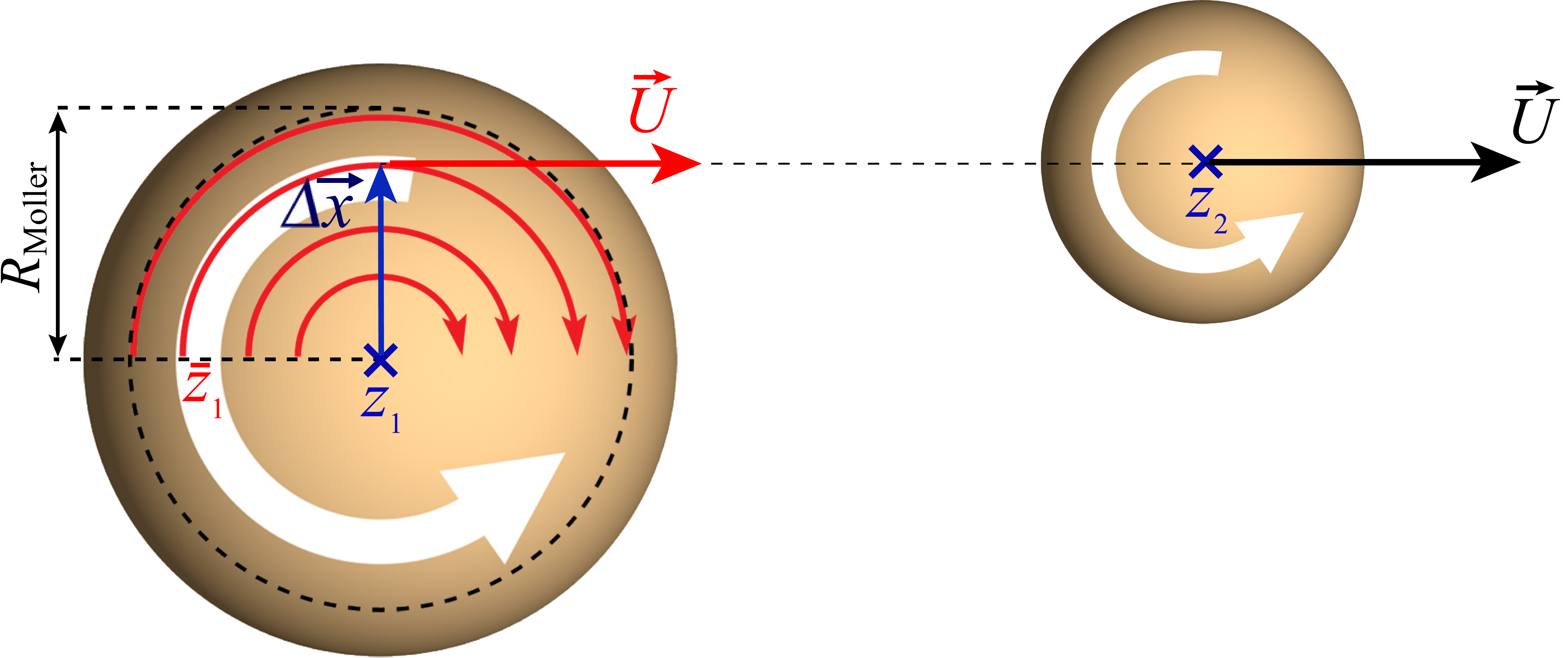}

\protect\caption{\label{fig:Determinacy}Two centroids, $\bar{z}_{1}^{\alpha}$ and
$z_{2}^{\alpha}$, of two different bodies for which the initial data
$\{z^{\alpha},m,U^{\alpha},S^{\alpha\beta}\}|_{{\rm in}}$ is the
same. The centroid $\bar{z}_{1}^{\alpha}$ is a \emph{helical} motion
of body 1, whose bulk (i.e., its nonhelical centroid $z_{1}^{\alpha}$)
is at rest; $z_{2}^{\alpha}$ is a \emph{nonhelical} motion of body
2, which moves uniformly with velocity $\vec{v}$. The figure corresponds
to flat spacetime, where the nonhelical centroids coincide with the
centroids as measured in the zero 3-momentum frames (defined by the
TD condition $S^{\alpha\beta}P_{\beta}=0$). Body 2 has a smaller
``intrinsic'' spin, but is more massive: $S_{\star2}=S_{\star1}/\gamma$,
$\mathcal{M}_{2}=\gamma\mathcal{M}_{1}$, $\gamma=1/\sqrt{1-v^{2}}$.
By giving the initial shift relative to the nonhelical centroid, $\Delta x^{\alpha}=S_{\ \beta}^{\alpha}P^{\beta}/\mathcal{M}^{2}$,
or, equivalently, the initial acceleration $a^{\alpha}=-\Delta x^{\alpha}\mathcal{M}^{2}/S^{2}$,
the degeneracy is removed. This is why the initial data $\{z^{\alpha},P^{\alpha},S^{\alpha\beta}\}|_{{\rm in}}$
(equivalent to $\{z^{\alpha},m,U^{\alpha},S^{\alpha\beta},a^{\alpha}\}|_{{\rm in}}$)
uniquely fixes the solution, and a definite velocity-momentum relation
$U^{\alpha}=U^{\alpha}(P^{\mu},S^{\mu\nu})$ exists.}
\end{figure}

The conflict between the two perspectives is only seeming. Given a
test body, with matter distribution described by some energy-momentum
tensor $T^{\alpha\beta}$, the condition $S^{\alpha\beta}U_{\beta}=0$
does not indeed specify a unique centroid; the MP SSC is obeyed by
an infinite set of worldlines. In the simplest case of flat spacetime,
as shown in \cite{MollerAIP,Helical}, every point within the so-called
``disk of centroids,''\footnote{In the zero 3-momentum frame (i.e., in the rest space orthogonal to
$P^{\alpha}$), the set of all possible positions of the center of
mass as measured by the different observers spans a disk, orthogonal
to $S_{\star}^{\alpha}$, of radius (\ref{eq:Rmoller}), centered
at the centroid as measured in that frame (TD centroid), see Fig.
\ref{fig:Determinacy}. Such disk is dubbed the ``disk of centroids.''
For more details see e.g. \cite{MollerAIP,CostaNatario2014}.} \emph{counterrotating} relative to the body with a certain fixed
angular velocity ($\OmHel=\mathcal{M}/S_{\star}$, where $S_{\star}$
is the angular momentum about the nonhelical centroid), yields a worldline
obeying the condition $S^{\alpha\beta}U_{\beta}=0$; this is depicted
in body 1 of Fig.~\ref{fig:Determinacy} (red semicircles therein).

The impact of this degeneracy in the initial value problem for the
equations of motion is not trivial though. This is why we devote the
rest of this section to explain how this difference between the MP
and other SSCs is reflected in the initial data set needed to determine
the solution. For all the SSCs apart from the MP one, one can apply
the initial data set $\{z^{\alpha},P^{\alpha},S^{\alpha\beta}\}|_{{\rm in}}$
equally well as the set $\{z^{\alpha},m,U^{\alpha},S^{\alpha\beta}\}|_{{\rm in}}$
, since both fix the solution uniquely. In the case of the MP SSC,
however, only the former data set provides a unique solution, whereas
the latter has to be supplemented by the initial acceleration, i.e.,
one needs the data set $\{z^{\alpha},m,U^{\alpha},S^{\alpha\beta},a^{\alpha}\}|_{{\rm in}}$.
The reason for this can be seen from the generic $P-U$ relation (\ref{eq:Momentum}).
Namely, for all the usual SSCs except from the MP one, it can be shown
that ${\rm D}V_{\beta}/{\rm d}\tau$ is a function of $(z^{\mu},U^{\mu},S^{\mu\nu})$,
which allows to obtain $P^{\alpha}$ as a function of $(z^{\mu},U^{\mu},S^{\mu\nu},m)$,
thereby rendering the set $\{z^{\alpha},P^{\alpha},S^{\alpha\beta}\}|_{{\rm in}}$
equivalent to $\{z^{\alpha},m,U^{\alpha},S^{\alpha\beta}\}|_{{\rm in}}$.
Below we provide the proof of the above statement.

In the case of the TD SSC, $V_{\beta}=P_{\beta}/\mathcal{M}$, we
have $\mu=\mathcal{M}$ and $\gamma(V,U)=m/\mathcal{M}$, therefore
Eq.~\eqref{Papa-p} gives that 
\[
\frac{{\rm D}V_{\beta}}{{\rm d}\tau}=\frac{1}{\mathcal{M}}\frac{{\rm D}P_{\beta}}{{\rm d}\tau}=-\frac{1}{2\mathcal{M}}R_{\beta\gamma\mu\nu}S^{\mu\nu}U^{\gamma}\ \quad.
\]
By contracting Eq.~\eqref{eq:Momentum} with $U^{\alpha}$ one obtains
\[
\mathcal{M}^{2}=m^{2}-\frac{1}{2}S^{\alpha\beta}U_{\alpha}R_{\beta\gamma\mu\nu}S^{\mu\nu}U^{\gamma}\ \quad,
\]
and by substituting the above $\mathcal{M}$ in \eqref{eq:Momentum}
one obtains the momentum $P^{\alpha}$ in terms of $(z^{\mu},U^{\mu},S^{\mu\nu},m)$.

In the case of the Corinaldesi-Papapetrou SSC \cite{Corinaldesi51},
$V^{\beta}=u_{{\rm lab}}^{\beta}$ is the congruence of ``laboratory''
observers \cite{CostaNatario2014}(at rest in the given coordinate
system). Thus, contracting Eq.~\eqref{eq:Momentum} with $U^{\alpha}$
leads to 
\[
\mu_{{\rm lab}}=m~\gamma(u_{{\rm lab}},U)+S_{\alpha\beta}u_{{\rm lab}}^{\beta;\sigma}U_{\sigma}U^{\alpha}\quad,
\]
where $\mu_{{\rm lab}}\equiv-P_{\alpha}u_{{\rm lab}}^{\alpha}$, $\gamma(u_{{\rm lab}},U)=-u_{{\rm lab}}^{\alpha}U_{\alpha}$,
and $u_{{\rm lab}}^{\alpha;\beta}$ is determined by the kinematics
of the observer congruence \cite{CostaNatario2014}. Substituting
into (\ref{eq:Momentum}) one obtains $P^{\alpha}$ in terms of $(z^{\mu},U^{\mu},S^{\mu\nu},m)$.

For the OKS condition, simply $P^{\alpha}=mU^{\alpha}$.

The case of the MP condition is different, as (\ref{eq:Momentum})
yields 
\[
P^{\alpha}=mU^{\alpha}+S^{\alpha\beta}\frac{{\rm D}U_{\beta}}{{\rm d}\tau}\ ;
\]
so clearly the initial values $\{z^{\alpha},m,U^{\alpha},S^{\alpha\beta}\}|_{{\rm in}}$
are not sufficient, since one cannot from them determine the acceleration
$a^{\alpha}={\rm D}U^{\alpha}/{\rm d}\tau$, which is needed in order
to obtain $P^{\alpha}$. Physically, this is because the same data
$\{z^{\alpha},m,U^{\alpha},S^{\alpha\beta}\}|_{{\rm in}}$ might correspond
to a nonhelical solution of a given physical body, as well as to helical
solutions of an indiscriminate number of physical bodies. This is
exemplified, for the case of flat spacetime, in Fig. \ref{fig:Determinacy}:
a given tensor $S^{\alpha\beta}$ and a 4-velocity $U^{\alpha}$ might
correspond to a \emph{helical} solution of body 1, whose bulk (i.e.,
its nonhelical centroid $z_{1}^{\alpha}$) is at rest, or to a \emph{nonhelical}
solution of body 2, which is uniformly moving with velocity $\vec{v}=\vec{U}/\gamma$.
This is so when their ``intrinsic'' spins (i.e., their angular momentum
about the nonhelical centroids $z_{1}^{\alpha}$ and $z_{2}^{\alpha}$)
and masses $\mathcal{M}=\sqrt{-P^{\alpha}P_{\alpha}}$ obey specific
relations that we shall now derive.

First notice, from Eq. (\ref{eq:Momentum}) applied to the Tulczyjew-Dixon
SSC ($V^{\alpha}=P^{\alpha}/\mathcal{M}$), that, in the absence of
forces (${\rm D}P^{\alpha}/{\rm d}\tau=0$), one has $P^{\alpha}=mU^{\alpha}$;
this implies that the TD centroid coincides with a centroid of the
MP SSC, more precisely the nonhelical one, since ${\rm D}P^{\alpha}/{\rm d}\tau=0\Rightarrow{\rm D}U^{\alpha}/{\rm d}\tau=0$
for such worldline. Therefore, the nonhelical centroids $z_{1}^{\alpha}$
and $z_{2}^{\alpha}$ are TD centroids. Now, recall the well known
flat spacetime expression (e.g. \cite{Gravitation}; see also Sec.
\ref{sec:Different-solutions-correspondin}) relating the angular
momentum tensors ($S^{\alpha\beta}$ and $\bar{S}^{\alpha\beta}$)
of a given body about two different points $z^{\alpha}$ and $\bar{z}^{\alpha}=z^{\alpha}+\Delta x^{\alpha}$,
\begin{equation}
\bar{S}^{\alpha\beta}=S^{\alpha\beta}+2P^{[\alpha}\Delta x^{\beta]}\ .\label{eq:SSbarflat}
\end{equation}
Let us moreover denote, as in \cite{Helical,CostaNatario2014}, by
$S_{\star}^{\alpha\beta}$ the angular momentum of a given body taken
about its TD centroid (so that $S_{\star}^{\alpha\beta}P_{\beta}=0$),
and by $S_{\star}^{\alpha}=-\epsilon_{\ \beta\gamma\delta}^{\alpha}S_{\star}^{\gamma\delta}P^{\beta}/(2\mathcal{M})$
the corresponding spin vector. The condition that $S^{\alpha\beta}$
be simultaneously the angular momentum of body 1 about its \emph{helical}
centroid $\bar{z}_{1}^{\alpha}$, and the angular momentum of body
2 about its \emph{nonhelical} centroid $z_{2}^{\alpha}$ (the TD centroid
of body 2), implies, for body 2, $S_{\star2}^{\alpha\beta}=S^{\alpha\beta}\Rightarrow S_{\star2}^{\alpha}=S^{\alpha}$,
and, for body 1, cf. Eq. (\ref{eq:SSbarflat}), 
\begin{align}
 & S^{\alpha\beta}=S_{\star1}^{\alpha\beta}+2P_{1}^{[\alpha}\Delta x^{\beta]}\ ;\label{eq:SSstar}\\
 & \Delta x^{\alpha}=-S_{\star1}^{\alpha\beta}\frac{U_{\beta}}{m}\quad,\label{eq:ShiftFlat}
\end{align}
where Eq. (\ref{eq:ShiftFlat}) follows from contracting (\ref{eq:SSstar})
with $U_{\beta}$, and making $\Delta x^{\beta}U_{\beta}=0$. The
vector $\Delta x^{\alpha}=\bar{z}_{1}^{\alpha}-z_{1}^{\alpha}$ is
the ``shift'' of the centroid $\bar{z}_{1}^{\alpha}$ relative to
$z_{1}^{\alpha}$; it is a vector orthogonal to the worldlines of
both centroids, that yields their instantaneous spatial displacement
(as measured in the rest frames of either of them). It is the analogue
of the Newtonian displacement vector, as illustrated in Fig. \ref{fig:Determinacy}.
It follows that 
\begin{equation}
S^{\alpha}\equiv-\frac{1}{2}\epsilon_{\ \beta\gamma\delta}^{\alpha}U^{\delta}S^{\beta\gamma}=\gamma S_{\star1}^{\alpha}-\epsilon_{\ \beta\gamma\delta}^{\alpha}U^{\delta}P_{1}^{\beta}\Delta x^{\gamma}=\frac{S_{\star1}^{\alpha}}{\gamma}\ ,\label{eq:SSstarvector}
\end{equation}
where $\gamma\equiv-U_{\alpha}P_{1}^{\alpha}/\mathcal{M}_{1}=m/\mathcal{M}_{1}$
satisfies\footnote{The factor $\gamma$ can also be written as $\gamma=1/\sqrt{1-v^{2}}$,
where $\vec{v}$ is the velocity of the centroid $\bar{z}_{1}^{\alpha}$
relative to the zero 3-momentum frame (the reference frame depicted
in Fig. \ref{fig:Determinacy}, where $\vec{v}=\vec{U}/\gamma$),
or, equivalently, the velocity of $\bar{z}_{1}^{\alpha}$ with respect
to $z_{1}^{\alpha}$. } $\gamma>1$. In both the second and third equalities of (\ref{eq:SSstarvector})
we notice that $U_{\delta}S_{\star1}^{\delta}=0$. To obtain this
relation, one first notes that substituting (\ref{eq:SSstar}) into
$S_{\star1}^{\alpha}=-\epsilon_{\ \beta\gamma\delta}^{\alpha}S_{\star1}^{\gamma\delta}P_{1}^{\beta}/(2\mathcal{M}_{1})$,
and contracting with $U_{\alpha}$, yields $S_{\star1}^{\alpha}U_{\alpha}=-S_{\alpha}P_{1}^{\alpha}/\mathcal{M}_{1}$;
then one just has to note, from Eq. (\ref{eq:momentumVelMP}), that
$S_{\alpha}P_{1}^{\alpha}=0$. We thus see that the data $\{z^{\alpha},m,U^{\alpha},S^{\alpha\beta}\}|_{{\rm in}}$
is the same for bodies 1 and 2 provided that 
\[
S_{\star1}^{\alpha}=\gamma S_{\star2}^{\alpha}=\gamma S^{\alpha}
\]
(so body 1 has a larger intrinsic spin than body 2, $S_{\star1}=\gamma S_{\star2}>S_{\star2}$),
and 
\[
m=\gamma\mathcal{M}_{1}=\mathcal{M}_{2}
\]
(so body 2 is more massive than body 1: $\mathcal{M}_{1}<\mathcal{M}_{2}$).

Such degeneracy is removed by additionally fixing the initial acceleration
$a^{\alpha}|_{{\rm in}}$. In fact, the initial data $\{z^{\alpha},m,U^{\alpha},S^{\alpha\beta},a^{\alpha}\}|_{{\rm in}}$
and $\{z^{\alpha},P^{\alpha},S^{\alpha\beta}\}|_{{\rm in}}$ are equivalent
under this spin condition, since from the latter one immediately obtains
$U^{\alpha}|_{{\rm in}}$ via (\ref{eq:momentumVelMP}), and also
$a^{\alpha}|_{{\rm in}}$ via the explicit expression for the acceleration
(\ref{eq:ExplicitaFMP}).

The way these things play out is especially intuitive again in the
flat spacetime case in Fig. \ref{fig:Determinacy}: as shown in detail
in \cite{Helical,MollerAIP}, for a given body (body 1 in Fig. \ref{fig:Determinacy}),
the MPD system (\ref{Papa-p})-(\ref{eq:Spinevol}) supplemented by
the MP SSC is satisfied by an infinite set of worldlines which, as
viewed from the perspective of the body's zero 3-momentum frame (the
frame represented in Fig. \ref{fig:Determinacy}), consist of a set
of circular motions (red semicircles), of radius $R=\|\Delta x^{\alpha}\|$,
centered at the nonhelical centroid ($z_{1}^{\alpha}$ in Fig. \ref{fig:Determinacy}).
Since, as explained above, the latter coincides with the body's TD
centroid, let us denote it henceforth by $z^{\alpha}(P)$. In other
frames, the solutions consist of a combination of such circular motion
with a boost parallel to $\vec{P}$. If one is given just the initial
data $\{z^{\alpha},m,U^{\alpha},S^{\alpha\beta}\}|_{{\rm in}}$, as
explained above, one has no way of knowing to which kind of solution
(helical or nonhelical) of which kind of body it corresponds to (i.e.,
which are the defining moments $P^{\alpha}$ and $S_{\star}^{\alpha\beta}$,
whether its bulk at rest or moving, etc). This is exemplified by bodies
1 and 2 of Fig. \ref{fig:Determinacy}, for which such data is the
same. For the initial data $\{z^{\alpha},P^{\alpha},S^{\alpha\beta}\}|_{{\rm in}}$
the situation is very different: the momentum $P^{\alpha}$ tells
us immediately the 4-velocity of the nonhelical centroid: ${\rm d}z^{\alpha}(P)/{\rm d}\tau=P^{\alpha}/\mathcal{M}$;
$P^{\alpha}$ and $S^{\alpha\beta}$ combined give us the shift $\Delta x^{\alpha}=z^{\alpha}-z^{\alpha}(P)$
via the expression 
\begin{equation}
\Delta x^{\alpha}=\frac{S_{\ \beta}^{\alpha}P^{\beta}}{\mathcal{M}^{2}}\label{eq:ShiftP}
\end{equation}
which follows from contracting (\ref{eq:SSstar}) with $P_{\beta}$
(identifying $P_{1}^{\alpha}\rightarrow P^{\alpha}$, $S_{\star1}^{\alpha\beta}\rightarrow S_{\star}^{\alpha\beta}$
therein), and noting that $\Delta x^{\beta}P_{\beta}=0$. From this
one gets the coordinates of the TD centroid $z^{\alpha}(P)$. In other
words, as depicted in Fig. \ref{fig:Determinacy}, the vector $\Delta x^{\alpha}$
tells us whether the motion is helical or not, and \emph{which one
of the helices}. Alternatively, the same information is given by the
initial acceleration, since, from Eqs. (\ref{eq:ExplicitaFMP}) and
(\ref{eq:ShiftP}), $a^{\alpha}=-\Delta x^{\alpha}\mathcal{M}^{2}/S^{2}$.
Moreover, the angular momentum $S_{\star}^{\alpha\beta}$ about the
nonhelical centroid $z^{\alpha}(P)$ can be obtained from $\Delta x^{\alpha}$
and $S^{\alpha\beta}$ using, again, (\ref{eq:SSstar}). The motion
is then totally determined, because we know the center {[}$z^{\alpha}(P)$,
that is, $z_{1}^{\alpha}$ in Fig. \ref{fig:Determinacy}{]} and the
radius ($\Delta x^{\alpha}$) of the circular motion described by
$z^{\alpha}$ around $z{}^{\alpha}(P)$; and we know moreover its
angular velocity, which, as shown in \cite{Helical,MollerAIP}, is
the same for all helices and equal to $\vec{\OmHel}=-\mathcal{M}\vec{S}_{\star}/S_{\star}^{2}$.
In this way we get an intuitive picture of why the motion (and hence
$U^{\alpha}$) is completely and uniquely determined given the initial
data $\{z^{\alpha},P^{\alpha},S^{\alpha\beta}\}|_{{\rm in}}$, making
natural the existence of the momentum-velocity relation (\ref{eq:momentumVelMP}).

\section{Different solutions corresponding to the same physical body\label{sec:Different-solutions-correspondin}}

In this section we discuss the degeneracy of the MP SSC, and the description
of a given physical body through the different representative worldlines
obeying this spin condition. First of all one needs to establish what,
in the framework of a pole-dipole approximation, defines a physical
body. In a multipole expansion, the energy-momentum tensor $T^{\alpha\beta}$
and the charge current density 4-vector ($j^{\alpha}$) of an extended
body are represented by its multipole moments (see e.g. \cite{Dixon1964,Madore:1969}).
To pole-dipole order, and in the absence of an electromagnetic field,
the momentum $P^{\alpha}$ and the spin tensor $S^{\alpha\beta}$
are the only of such moments entering the equations of motion. Such
moments are taken with respect to a reference worldline $z^{\alpha}(\tau)$
and defined as integrals over a certain spacelike hypersurface. Different
methods have been proposed for precisely defining the moments in a
curved spacetime. Some of them are based on bitensors \cite{Dixon1964,Dixon1970I,Wald et al 2010},
while others employ an exponential map \cite{Madore:1969}. In the
latter case the moments take the form \cite{Madore:1969,CostaNatario2014}
\begin{eqnarray}
P^{\hat{\alpha}} & \equiv & \int_{\Sigma(z,V)}T^{\hat{\alpha}\hat{\beta}}d\Sigma_{\hat{\beta}}\ ,\label{eq:Pgeneral}\\
S^{\hat{\alpha}\hat{\beta}} & \equiv & 2\int_{\Sigma(z,V)}x^{[\hat{\alpha}}T^{\hat{\beta}]\hat{\gamma}}d\Sigma_{\hat{\gamma}}\ ,\label{eq:Sab}
\end{eqnarray}
in a system of Riemann \emph{normal} coordinates $\{x^{\hat{\alpha}}\}$
originating at $z^{\alpha}$. Here $\Sigma(z,V)$ is the spacelike
hypersurface generated by all geodesics orthogonal to the timelike
vector $V^{\alpha}$ at the point $z^{\alpha}$, $d\Sigma$ is the
3-volume element on $\Sigma(z,V)$, and $d\Sigma_{\gamma}\equiv-n_{\gamma}d\Sigma$,
where $n^{\alpha}$ is the unit vector normal to $\Sigma(z,V)$ (at
$z^{\alpha}$, $n^{\alpha}=V^{\alpha}$).

For a free particle in flat spacetime, the conservation equations
$T_{\ \ ;\beta}^{\alpha\beta}=0$, along with the existence of a maximal
number of Killing vectors, imply that both $P^{\alpha}$ and $S^{\alpha\beta}$
are independent of $\Sigma(z,V)$ (see, e.g., \cite{Gravitation}).
Thus, $S^{\alpha\beta}$ is just a function of the reference point
$z^{\alpha}$, and $P^{\alpha}$ is a constant vector independent
of the point. Hence, given $P^{\alpha}$ and $S^{\alpha\beta}$ about
a reference worldline $z^{\alpha}(\tau)$, the moments of the same
body relative to another reference worldline $\bar{z}^{\alpha}(\bar{\tau})$
are such that, in \emph{a global rectangular coordinate system}, the
components of $P^{\alpha}$ remain the same, and $S^{\alpha\beta}$
is transformed by the well-known expression (\ref{eq:SSbarflat}).

In curved spacetime the situation is more complicated because the
moments depend on the hypersurface of integration $\Sigma$ (which
in turn are not simply hyperplanes, as in flat spacetime), and a simple,
\emph{exact} relation between the moments $\{\bar{P}^{\alpha},\bar{S}^{\alpha\beta}\}$
taken with respect to $\bar{z}^{\alpha}$, $\Sigma(\bar{z},\bar{V})$,
and the moments $\{P^{\alpha},S^{\alpha\beta}\}$ evaluated with respect
to $z^{\alpha}$, $\Sigma(z,V)$, does not exist. However, it is still
possible to devise a simple set of transformation rules that, to a
very good approximation, allows us to obtain the moments taken about
$\bar{z}^{\alpha}$ from the knowledge of the moments about $z^{\alpha}$,
if the size of the test body is small compared to the scale of the
curvature. More precisely, the latter assumption holds when $\lambda=\|\mathbf{R}\|\rho^{2}\ll1$,
where $\|\mathbf{R}\|$ is the magnitude of the Riemann tensor and
$\rho$ is the radius of the body.

To obtain these transformation rules one starts by noticing that when
$\lambda\ll1$, then for any point $z^{\alpha}$ within the convex
hull of the body's worldtube, $P^{\alpha}$ and $S^{\alpha\beta}$
are independent of the argument $V^{\alpha}$ of $\Sigma(z,V)$. This
is explicitly shown in the Appendix of \cite{CostaNatario2014}. Now,
let $\{x^{\tilde{\alpha}}\}$ be a system of normal coordinates originating
from the point $\bar{z}^{\alpha}$. These coordinates can be chosen
such that 
\[
x^{\tilde{\alpha}}=x^{\hat{\alpha}}-\bar{z}^{\hat{\alpha}}+\mathcal{O}(\|\mathbf{R}\|\|x^{\hat{\alpha}}-\bar{z}^{\hat{\alpha}}\|^{2}\|\bar{z}^{\hat{\alpha}}-z^{\hat{\alpha}}\|)\ ,
\]
cf. Eq. (11.12) of \cite{Brewin}. Therefore, $x^{\tilde{\alpha}}\simeq x^{\hat{\alpha}}-\bar{z}^{\hat{\alpha}}$,
provided that $z^{\alpha}$ and $\bar{z}^{\alpha}$ are two points
within the body's convex hull\footnote{More precisely, within the intersection of the body's worldtube with
any spacelike hypersurface $\Sigma(z,V)$, that can be interpreted
as the rest space of some observer of 4-velocity $V^{\alpha}$.} (as is the case for two centroids) \emph{and} that the condition
$\lambda\ll1$ holds. Aligning the time axis of the coordinate system
$\{x^{\hat{\alpha}}\}$ with $V^{\alpha}$, $\partial_{\hat{0}}^{\alpha}|_{z}=V^{\alpha}$,
we can thus take $P^{\hat{\alpha}}$, $\bar{P}^{\tilde{\alpha}}$,
$S^{\hat{\alpha}\hat{\beta}}$, $\bar{S}^{\tilde{\alpha}\tilde{\beta}}$
as integrals over the \emph{same hypersurface} $x^{\hat{0}}=0$, which,
using (\ref{eq:Pgeneral})-(\ref{eq:Sab}), leads to 
\begin{equation}
\bar{P}^{\tilde{\alpha}}=P^{\hat{\alpha}};\qquad\bar{S}^{\tilde{\alpha}\tilde{\beta}}=\bar{S}^{\hat{\alpha}\hat{\beta}}=S^{\hat{\alpha}\hat{\beta}}+2P^{[\hat{\alpha}}\Delta x^{\hat{\beta}]}\ ,\label{eq:TransfromNcoor}
\end{equation}
where $\Delta x^{\hat{\alpha}}=\bar{z}^{\hat{\alpha}}-z^{\hat{\alpha}}\equiv\bar{z}^{\hat{\alpha}}$.
This yields a rule for transition between different representations
of the same body: they are such that, in a normal coordinate system
originating at $z^{\alpha}$, the components $P^{\hat{\alpha}}$ of
the momentum are the same at both points, and the components of the
angular momentum obey relation (\ref{eq:TransfromNcoor}). The setting
of normal coordinates is however laborious in practical situations.

A practical covariant approach to implement these rules can be devised
as follows. First one notes that, since the system $\{x^{\hat{\alpha}}\}$
is constructed from geodesics radiating out of $z^{\alpha}$, the
components $\Delta x^{\hat{\alpha}}=\bar{z}^{\hat{\alpha}}$ are identified
with the vector $\Delta x^{\mu}$ at $z^{\alpha}$, tangent to the
geodesic $c^{\alpha}(s)$ connecting $z^{\alpha}$ and $\bar{z}^{\alpha}$,
and whose length equals that of the geodesic segment. And the point
$\bar{z}^{\alpha}$ is, thus, the image by the exponential map of
$\Delta x^{\mu}$ \cite{CostaNatario2014} (see Fig. \ref{fig:Shift}):
\begin{align}
\bar{z}^{\alpha} & =\exp_{z}^{\alpha}(\Delta x)=e^{\Delta s~{\rm d}/{\rm d}s}c^{\alpha}(s)|_{s=s_{z}}\label{eq:barz1}\\
 & =z^{\alpha}+\dot{c}^{\alpha}(s_{z})\Delta s+\frac{1}{2}\ddot{c}^{\alpha}(s_{z})\Delta s^{2}+...\quad,\nonumber 
\end{align}
where $\Delta s\equiv s_{\bar{z}}-s_{z}$. Choosing $s$ as the proper
length of $c^{\alpha}$, we have $\Delta s=\|\Delta x^{\mu}\|$ and
\begin{equation}
\dot{c}^{\alpha}(s_{z})=\frac{\Delta x^{\alpha}}{\|\Delta x^{\mu}\|}\ \quad;\label{eq:Shiftvector}
\end{equation}
reading $\ddot{c}^{\alpha}$ from the geodesic equation $\ddot{c}^{\alpha}+\Gamma_{\beta\gamma}^{\alpha}\dot{c}^{\beta}\dot{c}^{\gamma}=0$,
we obtain 
\begin{equation}
\bar{z}^{\alpha}=z^{\alpha}+\Delta x^{\alpha}-\frac{1}{2}\Gamma_{\beta\gamma}^{\alpha}|_{z}\Delta x^{\beta}\Delta x^{\gamma}+...\ \quad.\label{eq:barz2}
\end{equation}
Now let $\bar{g}_{\ \alpha}^{\kappa}\equiv\bar{g}_{\ \alpha}^{\kappa}(\bar{z},z)$
denote the bitensor that parallel propagates tensors $\mathcal{A}^{\alpha_{1}...\alpha_{n}}$
from $z^{\alpha}$ to $\bar{z}^{\alpha}$ along $c^{\alpha}(s)$ \cite{Dixon1964,DeWittBrehme}:
\begin{equation}
\mathcal{A}^{\alpha_{1}...\alpha_{n}}|_{\bar{z}}=\bar{g}_{\ \beta_{1}}^{\alpha_{1}}...\bar{g}_{\ \beta_{n}}^{\alpha_{n}}\mathcal{A}^{\beta_{1}...\beta_{n}}|_{z}\ \quad.\label{eq:BitensorParallel}
\end{equation}
Using the parallel transport equation ${\rm d}\mathcal{A}^{\alpha_{1}...\alpha_{n}}/{\rm d}s=-\Gamma_{\beta\gamma}^{\alpha_{1}}\mathcal{A}^{\beta\alpha_{2}...\alpha_{n}}\dot{c}^{\gamma}-...-\Gamma_{\beta\gamma}^{\alpha_{n}}\mathcal{A}^{\alpha_{1}...\alpha_{n-1}\beta}\dot{c}^{\gamma}$,
this is 
\begin{align}
\mathcal{A}^{\alpha_{1}...\alpha_{n}}|_{\bar{z}} & =\mathcal{A}^{\alpha_{1}...\alpha_{n}}|_{z}-\int_{z}^{\bar{z}}\Gamma_{\beta\gamma}^{\alpha_{1}}(x)\mathcal{A}^{\beta\alpha_{2}...\alpha_{n}}dx^{\gamma}\nonumber \\
 & -...-\int_{z}^{\bar{z}}\Gamma_{\beta\gamma}^{\alpha_{n}}(x)\mathcal{A}^{\alpha_{1}...\alpha_{n-1}\beta}dx^{\gamma}\ .\label{eq:ParallelTransport}
\end{align}
Noting that, in the normal coordinate system $\{x^{\hat{\alpha}}\}$,
$\|\Gamma_{\hat{\beta}\hat{\gamma}}^{\hat{\alpha}}(x)\|\sim\|\mathbf{R}\|\|x\|$,
it follows that $\mathcal{A}^{\hat{\alpha}_{1}...\hat{\alpha}_{n}}|_{\bar{z}}=\mathcal{A}^{\hat{\alpha}_{1}...\hat{\alpha}_{n}}|_{z}+\mathcal{O}(\|\bm{\mathcal{A}}\|\|\mathbf{R}\|\|\Delta x^{\mu}\|^{2})$.
Therefore, under the assumption $\lambda\ll1$, $\mathcal{A}^{\hat{\alpha}_{1}...\hat{\alpha}_{n}}|_{\bar{z}}\simeq\mathcal{A}^{\hat{\alpha}_{1}...\hat{\alpha}_{n}}|_{z}$.
In other words, the condition that, under the assumption $\lambda\ll1$,
the components $\mathcal{A}^{\hat{\alpha}_{1}...\hat{\alpha}_{n}}|_{\bar{z}}$
of a tensor at $\bar{z}^{\alpha}$ equal those of a tensor $\mathcal{A}^{\hat{\alpha}_{1}...\hat{\alpha}_{n}}|_{z}$
at $z^{\alpha}$ in \emph{normal coordinates} originating from $z^{\alpha}$,
is equivalent to saying that $\mathcal{A}^{\alpha_{1}...\alpha_{n}}|_{\bar{z}}$
is obtained by parallel transporting the tensor $\mathcal{A}^{\alpha_{1}...\alpha_{n}}|_{z}$
from $z^{\alpha}$ to $\bar{z}^{\alpha}$ along $c^{\alpha}(s)$.

Thus, given two points $z^{\alpha}$ and $\bar{z}^{\alpha}$, or a
point $z^{\alpha}$ and a shift vector $\Delta x^{\alpha}$, we have
a covariant method for ``transforming'' and then ``transferring''
the moments from $z^{\alpha}$ to $\bar{z}^{\alpha}$. Namely, first
one has to transform the spin tensor using Eq.~\eqref{eq:SSbarflat},
in which $\Delta x^{\alpha}$ is a \emph{vector at} the point $z^{\alpha}$
{[}defined by Eq.~\eqref{eq:Shiftvector}{]}. Note that this is a
well-defined operation for tensors at $z^{\alpha}$: it yields a tensor
$\bar{S}^{\alpha\beta}|_{z}$ \emph{at} $z^{\alpha}$, one whose components
in the normal coordinates $\{x^{\hat{\alpha}}\}$ of $z^{\alpha}$
happen to equal the components $\bar{S}^{\tilde{\alpha}\tilde{\beta}}$
of the spin tensor about $\bar{z}^{\alpha}$ in the normal coordinates
$\{x^{\tilde{\alpha}}\}$ of $\bar{z}^{\alpha}$ {[}see Eq.~\eqref{eq:TransfromNcoor}{]}.
Then, one parallel transports $P^{\alpha}$ and $\bar{S}^{\alpha\beta}|_{z}$
to $\bar{z}^{\alpha}$, i.e. 
\begin{align}
\bar{S}^{\alpha\beta}|_{\bar{z}} & =\bar{g}_{\ \gamma}^{\alpha}\bar{g}_{\ \delta}^{\beta}\bar{S}^{\gamma\delta}|_{z}\nonumber \\
 & =\bar{g}_{\ \gamma}^{\alpha}\bar{g}_{\ \delta}^{\beta}(S^{\gamma\delta}+2P^{[\gamma}\Delta x^{\delta]})|_{z}\ \quad,\label{eq:TransferS}\\
P^{\alpha}|_{\bar{z}} & =\bar{g}_{\ \beta}^{\alpha}P^{\beta}|_{z}\quad.\label{eq:TransferP}
\end{align}

The above procedure can be used to shift between different representative
(centroid) worldlines of a given body. Usually one has a solution
$z^{\alpha}(\tau)$ corresponding to some spin condition $S^{\alpha\beta}V_{\beta}=0$,
and wishes to know how to shift to a worldline $\bar{z}^{\alpha}(\bar{\tau})$
specified by another SSC $\bar{S}^{\alpha\beta}\bar{V}_{\beta}=0$.
That can be done as follows. Starting from a point $z^{\alpha}$ along
the worldline $z^{\alpha}(\tau)$, a point $\bar{z}^{\alpha}$ of
the new worldline, such that the method above holds, is reached via
Eq. (\ref{eq:barz1}) by an appropriate shift vector $\Delta x^{\alpha}$.
The vector $\Delta x^{\alpha}$ is obtained in turn as follows. One
prescribes a vector $\bar{V}^{\alpha}|_{z}$ at $z^{\alpha}$ (understood
to result from the parallel transport of the actual $\bar{V}^{\alpha}\equiv\bar{V}^{\alpha}|_{\bar{z}}$,
at the yet to be determined $\bar{z}^{\alpha}$, to $z^{\alpha}$,
i.e. $\bar{V}^{\alpha}=\bar{g}_{\ \beta}^{\alpha}\bar{V}^{\beta}|_{z}$);
then 
\begin{equation}
\Delta x^{\alpha}=-\frac{S^{\alpha\beta}\bar{V}_{\beta}|_{z}}{\bar{\mu}}\ \quad,\label{eq:DeltaxCov}
\end{equation}
where $\bar{\mu}\equiv-P^{\alpha}\bar{V}_{\alpha}$. In order to derive
Eq. (\ref{eq:DeltaxCov}), one must recall, from \cite{DeWittBrehme},
some properties of the parallel propagator $\bar{g}_{\alpha\beta}$
in Eq. (\ref{eq:BitensorParallel}). Namely, this tensor is not symmetric:
its second slot parallel transports vectors from $z^{\alpha}$ to
$\bar{z}^{\alpha}$, as indicated in Eq. (\ref{eq:BitensorParallel}),
whereas the first slot does the inverse path. That is, let $\bar{g}_{\alpha\beta}(z,\bar{z})$
be the bitensor whose second slot parallel transports tensors from
$\bar{z}^{\alpha}$ to $z^{\alpha}$ {[}i.e., the reciprocal of the
tensor $\bar{g}_{\alpha\beta}(\bar{z},z)\equiv\bar{g}_{\alpha\beta}$
in Eq. (\ref{eq:BitensorParallel}){]}; we have 
\begin{equation}
\bar{g}_{\alpha\beta}(z,\bar{z})=\bar{g}_{\beta\alpha}(\bar{z},z)\equiv\bar{g}_{\beta\alpha}\ ,\label{eq:parallelpropagator}
\end{equation}
cf. Eq. (1.36) of \cite{DeWittBrehme}. Now, contracting Eq. (\ref{eq:TransferS})
with $\bar{V}_{\beta}$, noting, from relation (\ref{eq:parallelpropagator}),
that $\bar{V}_{\beta}\bar{g}_{\ \delta}^{\beta}=\bar{V}_{\delta}|_{z}$,
and that, by definition, $\bar{S}^{\alpha\beta}|_{\bar{z}}\bar{V}_{\beta}=0$
and $\Delta x^{\delta}\bar{V}_{\delta}|_{z}=0$, one obtains Eq. (\ref{eq:DeltaxCov}).
The vector $\Delta x^{\alpha}$ is orthogonal to both $V^{\alpha}$
and $\bar{V}^{\alpha}|_{z}$; it yields, in the sense of the exponential
map, the instantaneous \emph{spatial }(with respect to either $V^{\alpha}$
or $\bar{V}^{\alpha}|_{z}$) displacement\footnote{This can readily be seen by aligning the time axis of the coordinate
system $\{x^{\hat{\alpha}}\}$ in Eq.~\eqref{eq:Sab} with $\bar{V}^{\alpha}|_{z}$,
i.e. $\partial_{\hat{0}}^{\alpha}|_{z}=\bar{V}^{\alpha}|_{z}$, leading
to 
\[
-S^{\hat{i}\hat{\alpha}}\bar{V}_{\hat{\alpha}}|_{z}=S^{\hat{i}\hat{0}}=\int_{\Sigma(z,\bar{V})}x^{\hat{i}}T^{\hat{0}\hat{\gamma}}d\Sigma_{\hat{\gamma}}\equiv\bar{\mu}\bar{z}^{i}\ .
\]
} of the centroid $\bar{z}^{\alpha}$ measured by an observer of 4-velocity
$\bar{V}^{\alpha}|_{z}$, relative to the centroid $z^{\alpha}$.
Equation (\ref{eq:DeltaxCov}), together with Eq.~\eqref{eq:barz1}
and Eqs.~\eqref{eq:TransferS}-\eqref{eq:TransferP}, provide all
the initial data needed to evolve the equations of motion~\eqref{Papa-p}-\eqref{eq:Spinevol},
\emph{provided} that they are coupled to a velocity-momentum relation
$U^{\alpha}=U^{\alpha}(z^{\mu},P^{\mu},S^{\mu\nu})$, thereby uniquely
determining the new worldline $\bar{z}^{\alpha}(\bar{\tau})$.

\subsection{Transition between different Mathisson-Pirani centroids}

\begin{figure}
\includegraphics[width=0.75\columnwidth]{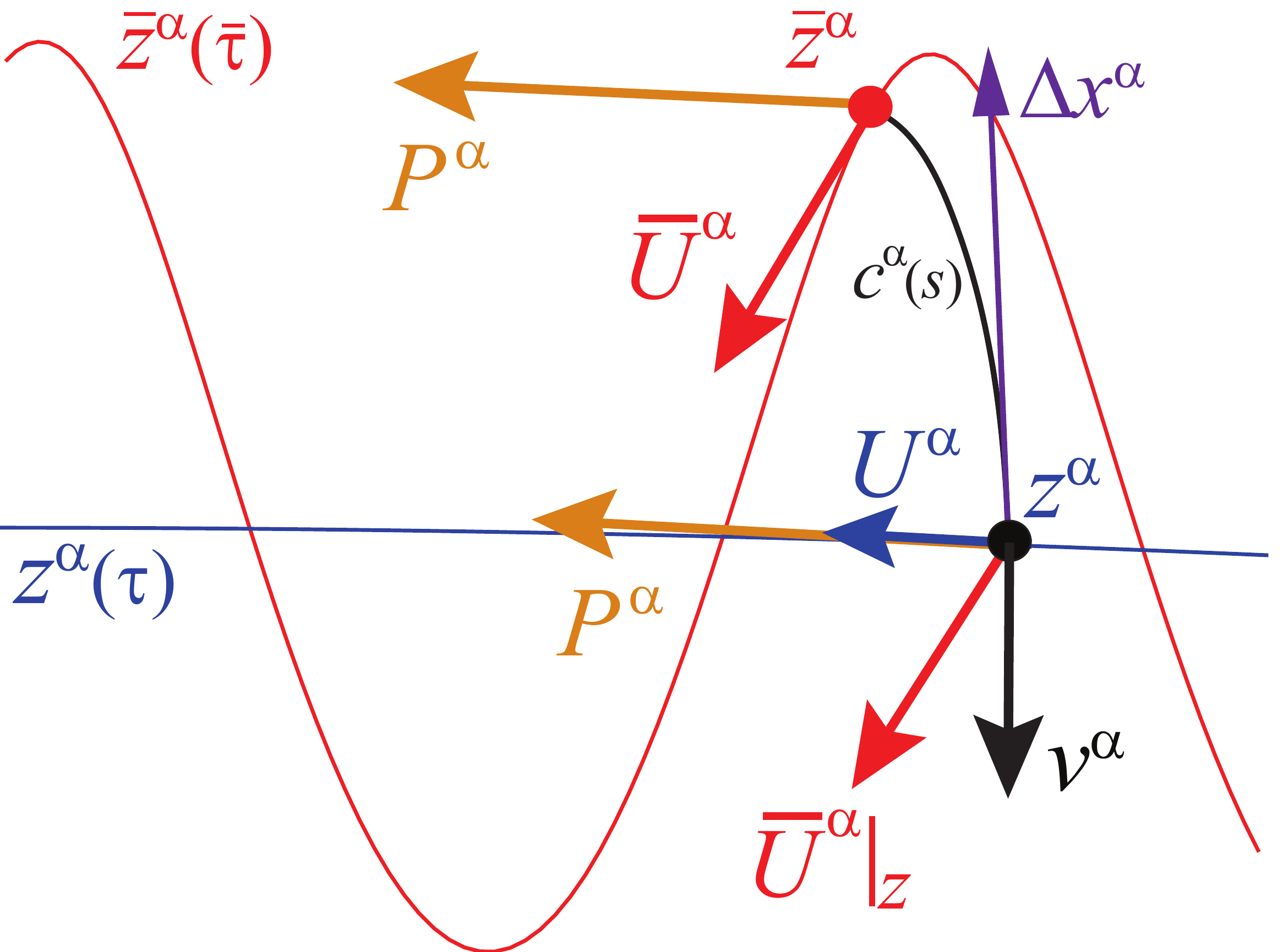}

\protect\caption{\label{fig:Shift}Two different centroids {[}of worldlines $\bar{z}^{\alpha}(\bar{\tau})$
and $z^{\alpha}(\tau)${]} corresponding to the same physical body.
The point $\bar{z}^{\alpha}=\exp_{z}^{\alpha}(\Delta x^{\mu})$ is
the image produced by the exponential map of the shift vector $\Delta x^{\alpha}=-S^{\alpha\beta}\bar{U}_{\beta}|_{z}/\bar{m}$
at $z^{\alpha}$, where $\bar{m}=-P^{\gamma}\bar{U}_{\gamma}|_{z}$.
$\bar{U}_{\beta}|_{z}$ is the vector resulting from parallel transporting
$\bar{U}^{\alpha}$ from $\bar{z}^{\alpha}$ to $z^{\alpha}$ along
the geodesic $c^{\alpha}(s)$ connecting these two points. $v^{\alpha}=\bar{U}^{\alpha}|_{z}/\gamma-U^{\alpha}$
is the ``kinematical'' relative velocity \cite{BolosIntrinsic}
of $\bar{z}^{\alpha}$ with respect to $z^{\alpha}$ (i.e. of $\bar{U}^{\alpha}$
with respect to $U^{\alpha}$). Given a worldline $z^{\alpha}(\tau)$
and its corresponding moments ($P^{\alpha}$ and $S^{\alpha\beta}$),
a new solution $\bar{z}^{\alpha}(\bar{\tau})$ of the MP SSC is completely
set up by prescribing, at some point $z^{\alpha}$, the vector $v^{\alpha}$
obeying the constraint~\eqref{eq:vcondition}, and then by using
Eqs.~\eqref{eq:barz1},~\eqref{eq:TransferS}-\eqref{eq:DeltaxCov}
for $\bar{V}^{\alpha}|_{z}=\bar{U}^{\alpha}|_{z}$.}
\end{figure}

According to the procedure above, given a solution $z^{\alpha}(\tau)$,
in order to change to a different worldline corresponding to a different
centroid of the same body, all one needs is prescribing the initial
vector $\bar{V}^{\alpha}|_{z}$ (i.e., the 4-velocity of the observer
with respect to which the new centroid is to be measured). All the
other quantities follow from Eqs.~\eqref{eq:barz1}, \eqref{eq:TransferS},
\eqref{eq:TransferP}, \eqref{eq:DeltaxCov}, that is: the new initial
position, spin vector, momentum, and shift vector, respectively.

The MP SSC demands $\bar{V}^{\alpha}$ to be tangent to the centroid
worldline, i.e. $\bar{V}^{\alpha}=\bar{U}^{\alpha}\equiv d\bar{z}^{\alpha}/d\tau$.
This demand does not specify a unique worldline, as already discussed
in Sec.~\ref{sec:The-duality-between}; but still it restricts the
choice of the eligible $\bar{V}^{\alpha}|_{z}=\bar{U}^{\alpha}|_{z}$,
as we shall now see. The conditions that $\bar{U}^{\alpha}$ must
obey can be found from the velocity-momentum relation~\eqref{eq:momentumVelMP}
re-written in terms of barred quantities, that is 
\begin{equation}
\bar{m}\bar{U}^{\alpha}=P^{\alpha}+\frac{1}{\bar{S}^{2}}\bar{S}^{\alpha\mu}\bar{S}_{\mu\beta}P^{\beta}\ \quad.\label{eq:Momentumvel}
\end{equation}
First note that $\bar{S}^{\alpha\mu}\bar{S}_{\mu}^{\ \beta}=\bar{S}^{\alpha}\bar{S}^{\beta}-\bar{h}^{\alpha\beta}\bar{S}^{2}$,
where $\bar{h}_{\alpha\beta}\equiv g_{\alpha\beta}+\bar{U}_{\alpha}\bar{U}_{\beta}$
is the space projector orthogonal to $\bar{U}^{\alpha}$ and 
\begin{align}
\bar{S}^{\alpha}=-\epsilon^{\alpha\beta\gamma\delta}\bar{S}_{\gamma\delta}\bar{U}_{\beta}/2\quad\label{eq:Sbar}
\end{align}
is Eq.~\eqref{eq:Sten2Svec} in barred quantities, it follows that
\[
\bar{m}\bar{U}^{\alpha}=P^{\alpha}-\bar{h}_{\ \beta}^{\alpha}P^{\beta}+\frac{1}{S^{2}}\bar{S}^{\alpha}\bar{S}^{\beta}P_{\beta}\quad\Leftrightarrow\quad\bar{S}^{\beta}P_{\beta}=0,
\]
since $P^{\alpha}-\bar{h}_{\ \beta}^{\alpha}P^{\beta}=\bar{m}\bar{U}^{\alpha}$.
Thus, Eq.~\eqref{eq:Momentumvel} is reduced to the orthogonality
between $\bar{S}^{\beta}$ and $P^{\beta}$, confirming the condition
suggested in \cite{Semerak II} (p.~1928) through a different route.

To see what this orthogonality implies for $\bar{U}^{\alpha}$, we
note that by contracting Eq.~\eqref{eq:Sbar} with $P_{\alpha}$
one gets 
\begin{align}
\bar{S}^{\alpha}P_{\alpha} & =-\frac{1}{2}\epsilon^{\alpha\beta\gamma\delta}\bar{S}_{\gamma\delta}\bar{U}_{\beta}P_{\alpha}=0\quad.\label{eq:ConditionsUinA}
\end{align}
If $S_{\star}^{\gamma\delta}|_{\bar{z}}$ is the angular momentum
about the centroid $z^{\alpha}(P)$ measured in the zero 3-momentum
frame (the TD centroid), parallel transported to $\bar{z}^{\alpha}$,
using $\bar{S}^{\gamma\delta}=S_{\star}^{\gamma\delta}|_{\bar{z}}+2P^{[\gamma}\zeta^{\delta]}$,
where $\zeta^{\tilde{\alpha}}=\bar{z}^{\tilde{\alpha}}-z^{\tilde{\alpha}}(P)$
is the shift vector from $z^{\alpha}(P)$ to $\bar{z}^{\alpha}$,
Eq.~\eqref{eq:ConditionsUinA} gives 
\begin{align}
\frac{1}{2}\epsilon_{\ \beta\gamma\delta}^{\alpha}S_{\star}^{\gamma\delta}|_{\bar{z}}\bar{U}^{\beta}P_{\alpha}=0\quad\Leftrightarrow\quad S_{\star}^{\beta}|_{\bar{z}}\bar{U}_{\beta}=0\quad.\label{eq:ConditionsUin}
\end{align}
Thus, the restriction imposed on $\bar{U}^{\alpha}$ is that it has
to be orthogonal to the spin vector $S_{\star}^{\beta}|_{\bar{z}}$
of the TD solution. Now, using properties (\ref{eq:parallelpropagator})
and $\bar{g}_{\alpha}^{\ \beta}\bar{g}_{\ \beta}^{\gamma}=\delta_{\alpha}^{\ \gamma}$
(cf. Eq. (1.35) of \cite{DeWittBrehme}), we can write $S_{\star}^{\alpha}|_{\bar{z}}\bar{U}_{\alpha}\equiv S_{\star}^{\alpha}|_{\bar{z}}\bar{g}_{\alpha}^{\ \beta}\bar{g}_{\ \beta}^{\gamma}\bar{U}_{\gamma}=S_{\star}^{\alpha}|_{z}\bar{U}_{\alpha}|_{z}$,
where $\bar{U}^{\alpha}|_{z}=\bar{g}_{\beta}^{\ \alpha}\bar{U}^{\beta}$
is the vector obtained by parallel transporting $\bar{U}^{\alpha}$
from $\bar{z}^{\alpha}$ to $z^{\alpha}$. Therefore, the condition
$S_{\star}^{\beta}|_{\bar{z}}\bar{U}_{\beta}=0$ is equivalent to
$S_{\star}^{\beta}|_{z}\bar{U}_{\beta}|_{z}=0$ (this is just the
statement that parallel transport preserves angles).

Consider now the spatial vector $v^{\alpha}$ defined by (see Fig.
\ref{fig:Shift}) 
\begin{equation}
\bar{U}^{\alpha}|_{z}=\gamma(U^{\alpha}+v^{\alpha});\qquad\gamma=-U^{\alpha}\bar{U}_{\alpha}|_{z}=1/\sqrt{1-v_{\alpha}v^{\alpha}}\ .\label{eq:vrel}
\end{equation}
The vector $v^{\alpha}$ is the \emph{kinematical relative velocity}
of the centroid $\bar{z}^{\alpha}$ with respect to $z^{\alpha}$
--- a natural generalization of the concept of relative velocity for
the case of objects located at different points \cite{BolosIntrinsic}.
Since $S_{\star}^{\beta}|_{z}U_{\beta}=0$ (as condition (\ref{eq:ConditionsUin})
must hold for \emph{any} solution $\bar{z}^{\alpha}$), that, together
with $S_{\star}^{\beta}|_{z}\bar{U}_{\beta}|_{z}=0$, implies via
(\ref{eq:vrel}) that 
\begin{equation}
S_{\star}^{\beta}|_{z}v_{\beta}=0\ .\label{eq:vcondition}
\end{equation}
In other words, compatibility of the initial data with the MP SSC
amounts to the requirement that $\bar{z}^{\alpha}$ moves relative
to $z^{\alpha}$ in a direction orthogonal to $S_{\star}^{\beta}|_{z}$.
For a free particle in flat spacetime, as depicted in Fig. \ref{fig:Determinacy},
this amounts to moving in a direction orthogonal to the body's axis
of rotation.

\begin{framed}%
\emph{Algorithm for transition between MP centroids} 
\begin{enumerate}
\item choose the ``kinematical relative velocity'' $v^{\alpha}$ of the
new centroid $\bar{z}^{\alpha}$ with respect $z^{\alpha}$, such
that it obeys (\ref{eq:vcondition}); 
\item determine $\bar{U}^{\alpha}|_{z}\equiv\bar{V}^{\alpha}|_{z}$ and
the shift vector $\Delta x^{\alpha}$ through Eqs. (\ref{eq:vrel})
and (\ref{eq:DeltaxCov}). 
\item Determine $\bar{z}^{\alpha}$ from Eq. (\ref{eq:barz2}). 
\item Parallel transport $P^{\alpha}$ to $\bar{z}^{\alpha}$ using Eq.
(\ref{eq:TransferP}); transform the spin tensor and parallel transport
it to $\bar{z}^{\alpha}$ using Eq. (\ref{eq:TransferS}). 
\item Use the obtained $\{\bar{z}^{\alpha},P^{\alpha},\bar{S}^{\alpha\beta}\}$
as initial data for the system (\ref{eq:EqsMotion}), uniquely determining
the solution.\end{enumerate}
\end{framed}

\section{Examples\label{sec:Examples}}

In this section we will employ the Mathisson-Pirani condition in physical
systems where it is easy to setup the nonhelical solution, and this
spin condition is especially suitable in that it leads to the simplest
description of the physical motion. In each case we will also exemplify
the helical descriptions of the same (within the realm of the pole-dipole
approximation) physical motion.

\subsection{Radial fall in Schwarzschild spacetime\label{sub:Radial-fall-in}}

\begin{figure}
\centering{}\includegraphics[width=1\columnwidth]{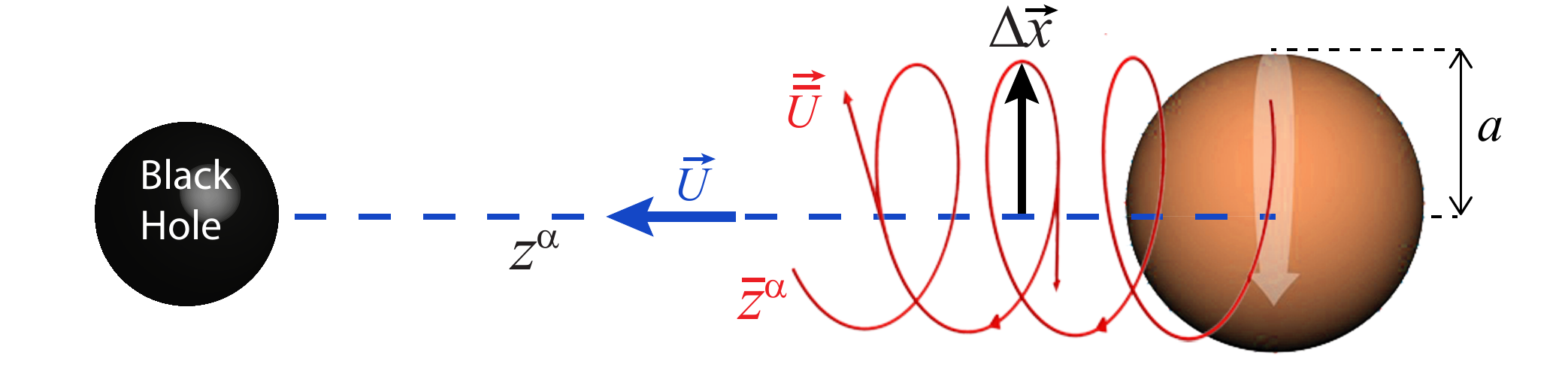} {\includegraphics[width=0.52\columnwidth]{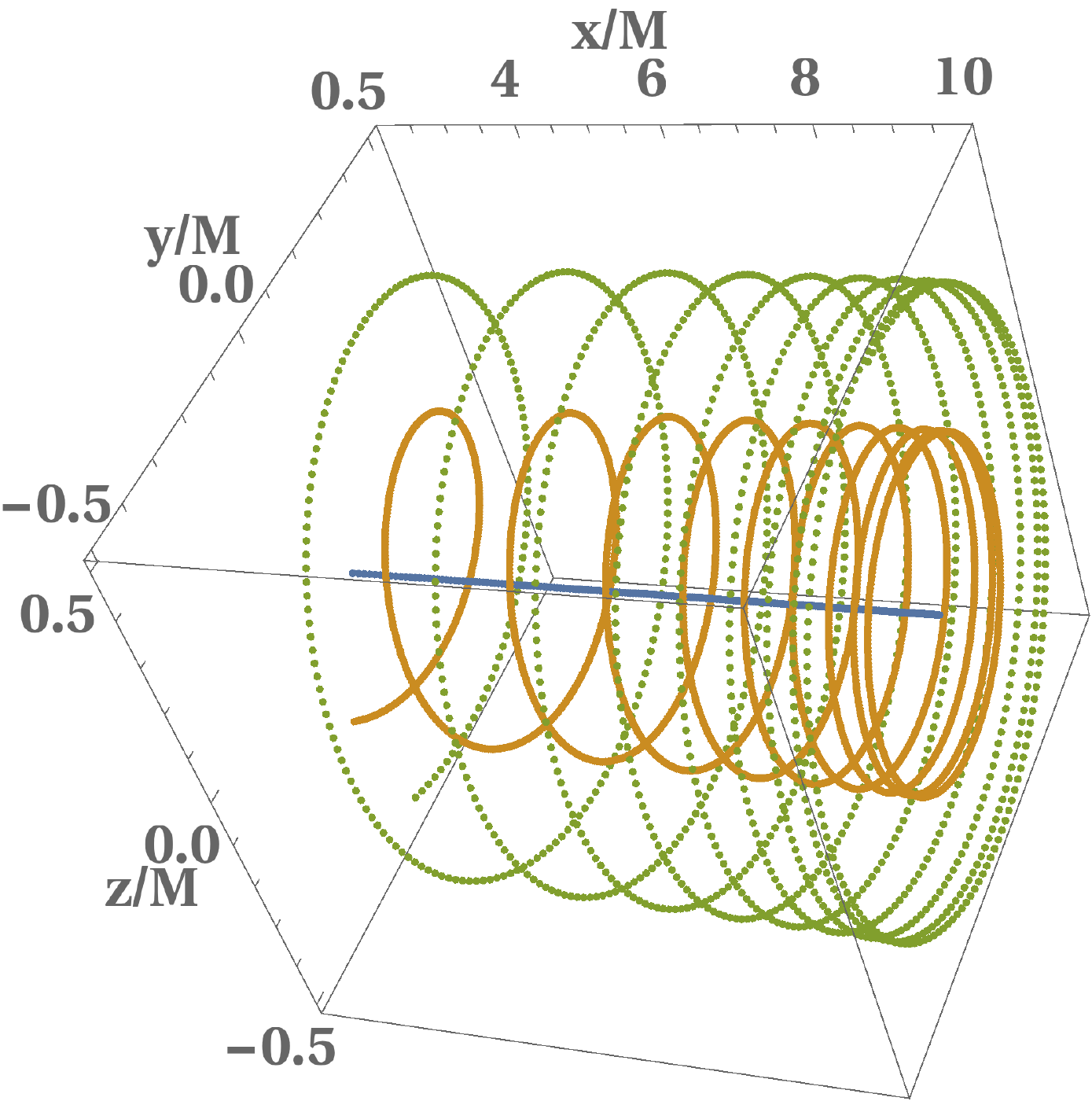}~~\includegraphics[width=0.46\columnwidth]{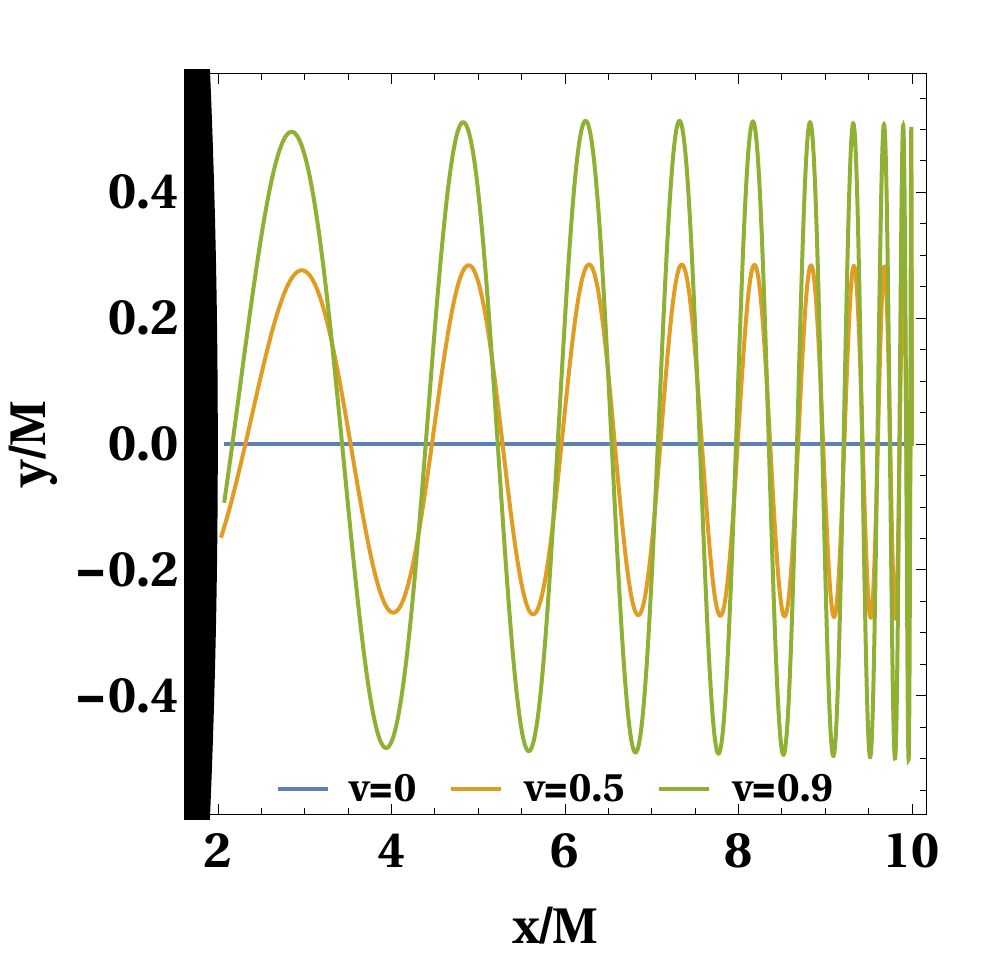}}
\protect\caption{\label{fig:RadialSpin} Left bottom panel: three different solutions---a
nonhelical centroid (blue straight line), plus two helical ones---of
the Mathisson-Pirani SSC, all representing, initially, \emph{the same
physical situation}: a spinning body with radial spin $S^{\alpha}=S^{r}\partial_{r}^{\alpha}$,
falling radially into the black hole (top panel). The nonhelical centroid
$z^{\alpha}(\tau)$ (hence the physical body) starts from rest at
$r=10M$: $U_{{\rm in}}^{\alpha}=U^{0}\partial_{0}^{\alpha}$; $S^{\alpha}$
(taken about $z^{\alpha}(\tau)$), has magnitude $S=0.5mM$. The helical
motions (which \emph{counterrotate} with the body) are prescribed
as having initial azimuthal velocity $v^{\alpha}=v^{\phi}\partial_{\phi}^{\alpha}$
relative to $z^{\alpha}(\tau)$, of magnitudes $v=0.5$ and $v=0.9$.
Their initial position is shifted from $z_{{\rm in}}^{\alpha}$ by
$\Delta\vec{x}|_{{\rm in}}=(1/\mathcal{M})(\vec{S}\times\vec{v})^{\theta}\partial_{\theta}$.
Right bottom panel: the corresponding 2D x-y plot (black region represents
the event horizon). The coordinates $\{x,y,z\}$ relate to Schwarzschild
coordinates by $x=r\sin\theta\cos\phi$, $y=r\sin\theta\sin\phi$,
$z=r\cos\theta$. }
\end{figure}

We wish to study the setup shown in Fig. \ref{fig:RadialSpin}, corresponding
to the motion of a physical body whose bulk has initial radial velocity
in the Schwarzschild spacetime. We start by setting an initial 4-velocity
$U^{\alpha}=U^{0}\partial_{t}^{\alpha}+U^{r}\partial_{r}^{\alpha}$.
For such $U^{\alpha}$, $\mathbb{H}_{\alpha\beta}=0$ (cf. Eq. (50)
of \cite{CostaNatarioZilhao}), and so the spin-curvature force (\ref{eq:TidaltensorF})
is zero (regardless of the orientation of $S^{\alpha}$). Taking this
into account, and using Eqs. (\ref{p,u,dSu}), (\ref{eq:Sten2Svec}),
leads to the equation of motion for the centroid 
\begin{equation}
\frac{{\rm D}P^{\alpha}}{{\rm d}\tau}=0~\Leftrightarrow~ma^{\alpha}+\epsilon_{\ \beta\gamma\delta}^{\alpha}U^{\delta}\frac{{\rm D}(S^{\beta}a^{\gamma})}{{\rm d}\tau}=0\ \quad.\label{eq:Force Ker-dS}
\end{equation}
This equation admits the trivial solution $a^{\alpha}=0$, which corresponds
to a radial geodesic trajectory. This solution, call it $z^{\alpha}(\tau)$,
is (obviously) the nonhelical MP centroid of this physical system,
and it is the same for any spinning body regardless the orientation
of its spin. For this special, geodesic case, Eq.~(\ref{p,u,dSu})
yields $P^{\alpha}=mU^{\alpha}$, which in turn implies that $z^{\alpha}(\tau)$
coincides with the (unique) centroid given by the TD SSC, i.e., it
holds that $S^{\alpha\beta}P_{\beta}\equiv S_{\star}^{\alpha\beta}P_{\beta}=0$.

Let us briefly discuss the description of the same physical motion
through other spin conditions. First notice that it is only under
the MP condition that the spin-curvature force takes the tidal tensor
form (\ref{eq:TidaltensorF}), which depends only on the centroid's
4-velocity $U^{\alpha}$ and on the spin vector $S^{\alpha}$; for
other SSCs the force (\ref{Papa-p}) depends also on $V^{\alpha}$,
as manifest in Eq. (\ref{eq:ForceSGen}). Starting with the TD condition,
$V^{\alpha}=P^{\alpha}/\mathcal{M}$, the motion cannot be set up
by prescribing a radial $U^{\alpha}$, for it is not possible to obtain
$P^{\alpha}$ from either the $U-P$ relation (\ref{eq:P_U_Dixon}),
nor the $P-U$ relation (\ref{eq:Momentum}). The problem is solved
instead by prescribing a radial momentum $P^{\alpha}=P^{0}\partial_{0}^{\alpha}+P^{r}\partial_{r}^{\alpha}$.
Then, by noticing that the numerator of the second term of Eq. (\ref{eq:P_U_Dixon})
can be written as $-4S^{\mu\nu}(\mathbb{H}^{P})_{\beta\nu}S^{\beta}$,
where $(\mathbb{H}^{P})_{\alpha\gamma}=\star R_{\alpha\beta\gamma\delta}P^{\beta}P^{\delta}/\mathcal{M}^{2}$,
and that, for a radial $P^{\alpha}$, $(\mathbb{H}^{P})_{\alpha\gamma}=0$
(cf. Eq. (50) of \cite{CostaNatarioZilhao}), we see that indeed Eq.
(\ref{eq:P_U_Dixon}) yields $P^{\alpha}=mU^{\alpha}$, leading to
the same solution obtained with the MP condition. Since such solution
is a radial geodesic, it obviously coincides as well with a particular
solution of the OKS condition, with $V^{\alpha}=U^{\alpha}$. Under
other spin conditions the situation is however more complicated; the
centroids that correspond to a body \emph{whose bulk} falls radially,
are, in general, shifted relative to the common centroid of the MP,
TD, and OKS conditions, and do \emph{not} move radially. Both the
spin-curvature force and the derivative of the hidden momentum are
in general \emph{not} zero for such centroids, leading to a nonzero
acceleration. This is the case of the centroids specified by the Corinaldesi-Papapetrou
and Newton-Wigner SSC's, which deflect as the body approaches the
black hole, cf. Eqs. (4.1), (4.2), (5.1) and (5.4) of \cite{BakerOConnel19741975},
and Fig. 6(c) of \cite{CostaNatario2014}. It is also the case of
the ``eccentric'' centroids of the OKS SSCs, which move nearly parallel
to the radial geodesic, see Eq. (45) and Fig. 6(d) of \cite{CostaNatario2014}
(in this case the hidden momentum is zero, the accelerated ``parallel''
trajectory being ensured by the spin-curvature force). Conversely,
if one naively prescribes an initial radial velocity $\vec{U}$ for
such centroids, then the solution will \emph{not} correspond to a
radial motion, but to another physical motion where the body's bulk
does not move radially. Therefore, without the knowledge of the radial
geodesic solution (obtained either with the MP, TD or OKS conditions),
it would not be clear whether a radial motion of the body's bulk occurs,
and how to prescribe the initial conditions for the corresponding
centroids.

\subsubsection{Helical centroids}

\begin{figure}
\includegraphics[width=1\columnwidth]{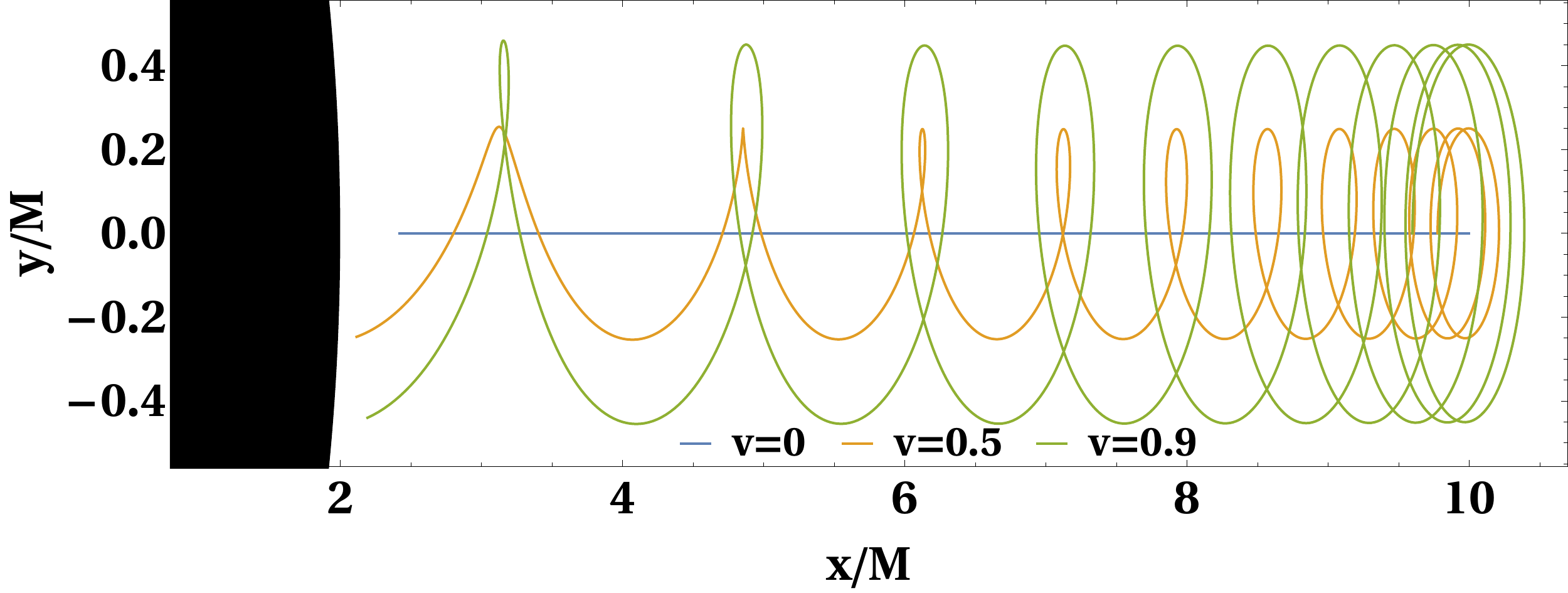} \protect\caption{\label{fig:PolarSpin}The analogue of Fig.~\ref{fig:RadialSpin}
for the case that the body's spin is aligned with the polar axis,
$S^{\alpha}=S^{\theta}\partial_{\theta}^{\alpha}$. The spin vector
about the nonhelical centroid (blue line) has again magnitude $S=0.5mM$,
and the ``helical'' motions are prescribed by putting initially
$v^{\alpha}=v^{\phi}\partial_{\phi}^{\alpha}$, of magnitudes $v=0.5$
and $v=0.9$. Their initial position is radially shifted from $z_{{\rm in}}^{\alpha}$
by $\Delta\vec{x}|_{{\rm in}}=(1/\mathcal{M})(\vec{S}\times\vec{v})^{r}\partial_{r}$.
Unlike the situation for radial spin in Fig. \ref{fig:PolarSpin},
here the ``helices'' are planar motions, lying in the equatorial
plane $\theta=\pi/2$.}
\end{figure}

To study the helical solutions, we consider two special cases: 
\begin{itemize}
\item radial spin $S^{\alpha}=S^{r}\partial_{r}^{\alpha}$ (Fig. \ref{fig:RadialSpin}), 
\item polar spin $S^{\alpha}=S^{\theta}\partial_{\theta}^{\alpha}$ (Fig.
\ref{fig:PolarSpin}). 
\end{itemize}
In both cases we consider that the body's bulk, which in this case
is faithfully represented by its nonhelical centroid $z^{\alpha}(\tau)$,
starts from rest. Since this baseline coincides with the TD centroid,
the shift equation~\eqref{eq:DeltaxCov} reduces to 
\begin{equation}
\Delta x^{\alpha}=-\frac{S^{\alpha\beta}v_{\beta}}{\mathcal{M}}\equiv-\frac{S_{\star}^{\alpha\beta}v_{\beta}}{\mathcal{M}}\ \quad.\label{eq:ShiftRadial}
\end{equation}
Moreover, since $z^{\alpha}(\tau)$ starts from rest, the initial
kinematical relative velocity $v^{\alpha}$ of the centroid $\bar{z}^{\alpha}(\bar{\tau})$
with respect to $z^{\alpha}(\tau)$ coincides with the velocity with
respect to the static observers. In both cases, we choose azimuthal
\emph{initial} velocities: $v^{\alpha}=v^{\phi}\partial_{\phi}^{\alpha}$,
leading to shift vectors along $\partial_{\theta}$ and $\partial_{r}$
for the radial and polar spin cases, respectively. We approximate
the initial position $\bar{z}^{\alpha}|_{{\rm in}}$ of the helical
centroids as shown in Eq. (\ref{eq:barz2}); we also expand in Taylor
series about $z^{\alpha}$ the Christoffel symbols in Eq. (\ref{eq:ParallelTransport}),
keeping only the lowest order term: $\Gamma_{\beta\gamma}^{\alpha}(x)=\Gamma_{\beta\gamma}^{\alpha}(z)+\mathcal{O}(x-z)$.
Thus, the expressions (\ref{eq:TransferP}) for the moments parallel
transported to $\bar{z}^{\alpha}$ are approximated by 
\begin{align}
P^{\alpha}|_{\bar{z}} & =\bar{g}_{\ \beta}^{\alpha}P^{\beta}|_{z}\simeq P^{\alpha}|_{z}-\Gamma_{\beta\gamma}^{\alpha}(z)P^{\beta}|_{z}\Delta x^{\gamma}\ \quad,\label{eq:PtransferAprox}\\
\bar{S}^{\alpha\beta}|_{\bar{z}} & \simeq\bar{S}^{\alpha\beta}|_{z}-\Gamma_{\delta\gamma}^{\alpha}\bar{S}^{\delta\beta}|_{z}\Delta x^{\gamma}-\Gamma_{\delta\gamma}^{\beta}\bar{S}^{\alpha\delta}|_{z}\Delta x^{\gamma}\ ~.\label{eq:STransferAprox}
\end{align}
This provides initial data for the helical solutions, which are then
numerically evolved using the equations of motion~\eqref{Papa-p}-\eqref{eq:Spinevol},
together with the momentum-velocity relation~\eqref{eq:momentumVelMP}.
We obtain two types of ``helical'' motion. On one hand, in the radial
spin case of Fig.~\ref{fig:RadialSpin}, they are proper helices,
winding about the (geodesic) nonhelical trajectory. On the other hand,
in the polar spin case, Fig.~\ref{fig:PolarSpin}, the result is
quite close to a superposition of an infalling radial geodesic (the
nonhelical solution) with a circular motion on the $\theta=\pi/2$
plane. The fact that, for both trajectories, the winding is about
the nonhelical centroid was to be expected from the fact that such
centroid coincides with the TD centroid, which is the center of the
disk of the possible centroids (see \cite{CostaNatario2014}). Indeed,
Eq. (\ref{eq:ShiftRadial}), whose space part, in the zero 3-momentum
frame and in vector notation, reads 
\begin{equation}
\Delta\vec{x}=\frac{\vec{S}_{\star}\times\vec{v}}{\mathcal{M}}\ ,\label{eq:ShiftHelical}
\end{equation}
tells us that the shift vectors corresponding to all the possible
helical solutions span a disk orthogonal to both $P^{\alpha}$ and
$S_{\star}^{\alpha}$, of radius (``Møller radius'') 
\begin{equation}
R_{{\rm Moller}}=\frac{S_{\star}}{\mathcal{M}}\ ,\label{eq:Rmoller}
\end{equation}
in the tangent space at $z^{\alpha}$. Such a situation resembles
the behavior of a free particle in flat spacetime (see \cite{Supplement2},
Sec. 1), only now the winding stretches for decreasing $r$ (unlike
in flat spacetime) due to the increase in radial velocity caused by
the black hole's gravitational field. The plots also indicate that
(contrary to the flat spacetime case) the amplitude of the helices
is not constant. In particular, as the particle approaches the horizon
the amplitude slightly decreases in the radial spin case (Fig.~\ref{fig:RadialSpin}),
whilst it slightly increases in the case of polar spin (Fig.~\ref{fig:PolarSpin}).
The amplitude changes are however very slight in both cases.

Let us stress, however, an important difference in the dynamics comparing
to the flat spacetime case: in flat spacetime, no force is exerted
on any of the centroids ($F^{\alpha}=0$), the helical-motion acceleration
comes only from an interchange between the kinetic momentum ($mU^{\alpha}$)
and the hidden momentum $\Phid^{\alpha}$ (see Fig.~3 of \cite{Helical}).
Here, however, the spin-curvature force (\ref{eq:TidaltensorF}) is
nonzero along all helical trajectories ($F^{\alpha}\ne0$). Thus the
acceleration results from the combined effects of the force and the
hidden momentum variation. The role of the force, however, is actually
to prevent the worldlines from diverging/converging, counteracting
the tidal forces due to the curvature, and ensuring that, from the
point of view of the zero 3-momentum frame, the helical motions stay
close to what they would be in flat spacetime. This is what we are
going to show next.

\subsubsection{An analysis of the helical dynamics through the deviation of worldlines\label{sub:WorldlineDev}}

\label{sec:DevEq}

We start by noticing that since, as discussed in Sec. \ref{sec:The-duality-between},
every point within the worldtube of centroids coincides momentarily
with a certain unique helical centroid, the helical solutions form
locally a \emph{congruence} of worldlines filling the worldtube of
centroids. Let $U^{\alpha}$ be the unit vector field tangent to such
a congruence of worldlines, and $\delta x^{\alpha}$ be a connecting
vector between different worldlines, so that it is Lie dragged along
the congruence, $\LieU\delta x^{\alpha}=0$. The latter expression
implies that ${\rm D}^{2}\delta x^{\alpha}/{\rm d}\tau^{2}\equiv\CovU\CovU\delta x^{\alpha}=\CovU\nabla_{\delta x}U^{\alpha}$,
and, thus, 
\begin{align}
\frac{{\rm D}^{2}\delta x^{\alpha}}{{\rm d}\tau^{2}} & =\nabla_{\delta x}a^{\alpha}-[\nabla_{\delta x},\CovU]U^{\alpha}\nonumber \\
 & =\nabla_{\delta x}a^{\alpha}-\mathbb{E}_{\ \gamma}^{\alpha}\delta x^{\gamma}\quad,\label{eq:Wdev1}
\end{align}
where $\mathbb{E}_{\alpha\beta}\equiv R_{\alpha\mu\beta\nu}U^{\mu}U^{\nu}$
is the electric part of the Riemann tensor. This is the deviation
equation for accelerated worldlines \cite{BiniStrains}, i.e. a generalization
of the geodesic deviation equation to nongeodesic curves. From the
relation $P_{{\rm hid}}^{\alpha}=P^{\alpha}-mU^{\alpha}$ {[}cf. Eq.
\eqref{eq:HidddenMom0}{]}, we have $a^{\alpha}=(F^{\alpha}-\CovU\Phid^{\alpha})/m$.
Substituting the latter into Eq.~\eqref{eq:Wdev1} leads to 
\begin{equation}
\frac{{\rm D}^{2}\delta x^{\alpha}}{{\rm d}\tau^{2}}=-\mathbb{E}_{\ \gamma}^{\alpha}\delta x^{\gamma}+\nabla_{\delta x}\frac{F^{\alpha}}{m}-\nabla_{\delta x}\CovU\frac{\Phid^{\alpha}}{m}\ \quad.\label{eq:Wdev12}
\end{equation}

To dipole order, the covariant derivative (along $\delta x^{\alpha}$)
of Eq.~\eqref{Papa-p} reads 
\begin{align}
\nabla_{\delta x}F^{\alpha}\simeq & -\frac{1}{2}R_{\ \beta\mu\nu}^{\alpha}U^{\beta}\nabla_{\delta x}S^{\mu\nu}\nonumber \\
 & -\frac{1}{2}R_{\ \beta\mu\nu}^{\alpha}S^{\mu\nu}\CovU\delta x^{\beta}\ \quad.\label{eq:DeltaF0}
\end{align}
The term $U^{\beta}S^{\mu\nu}\nabla_{\delta x}R_{\ \beta\mu\nu}^{\alpha}\equiv U^{\beta}R_{\ \beta\mu\nu;\lambda}^{\alpha}\delta x^{\lambda}S^{\mu\nu}$,
being of order $\mathcal{O}(\delta x^{\lambda}S^{\mu\nu})$, was neglected,
since $S^{\mu\nu}\delta x^{\lambda}\lesssim m\rho^{2}$, recall that
$\rho$ is the body's radius. The second term, however, is not negligible
to dipole order, since $\CovU\delta x^{\beta}=\nabla_{\delta x}U^{\beta}=U_{\ ;\alpha}^{\beta}\delta x^{\alpha}$
and $U_{\ ;\alpha}^{\beta}$ is of the order of the angular velocity
of the helical motions, $\OmHel=\mathcal{M}/S_{\star}=\mathcal{O}(S^{-1})$.
To compute $\nabla_{\delta x}S^{\mu\nu}$, it is convenient to use
the normal coordinate system $\{x^{\hat{\alpha}}\}$ originating from
$z^{\alpha}$, where the tensor function $S^{\hat{\mu}\hat{\nu}}(x)=S^{\hat{\mu}\hat{\nu}}(z)+2P^{[\hat{\mu}}x^{\hat{\nu}]}$
yields the angular momentum taken about any point of coordinates $x^{\hat{\alpha}}$
in terms of $x^{\hat{\alpha}}$ and the angular momentum about the
origin $S^{\hat{\mu}\hat{\nu}}(z)$. At the origin of such coordinates
one has therefore $S^{\hat{\mu}\hat{\nu}}(x)_{;\lambda}=S^{\hat{\mu}\hat{\nu}}(x)_{,\lambda}=2P^{[\hat{\mu}}\delta_{\hat{\lambda}}^{\hat{\nu}]}$;
the latter expression is however a tensor, so, in an arbitrary coordinate
system, we may write $S^{\mu\nu}(x)_{;\lambda}=2P^{[\mu}\delta_{\lambda}^{\nu]}$
and 
\begin{align}
\nabla_{\delta x}S^{\mu\nu} & =S^{\mu\nu}(x)_{;\lambda}\delta x^{\lambda}=2P^{[\mu}\delta_{\lambda}^{\nu]}\delta x^{\lambda}\nonumber \\
 & =2P^{[\mu}\delta x^{\nu]}\ \quad.\label{eq:SpinGrad}
\end{align}
We note in passing that Eq. (\ref{eq:SpinGrad}) actually holds for
an arbitrary infinitesimal displacement vector $\delta\mathcal{V}^{\alpha}$
(not necessarily the connecting vector $\delta x^{\alpha}$): $\nabla_{\delta\mathcal{V}}S^{\mu\nu}=2P^{[\mu}\delta\mathcal{V}^{\nu]}$;
if one takes $\delta\mathcal{V}^{\alpha}=d\tau U^{\alpha}$, we obtain
$\nabla_{U}S^{\mu\nu}=2P^{[\mu}U^{\nu]}$, and hence a very simple
derivation of the spin evolution equation~\eqref{eq:Spinevol}.

Taking, for simplicity, the nonhelical centroid as the basis worldline,
we have $\nabla_{\delta x}S^{\mu\nu}=2mU^{[\mu}\delta x^{\nu]}$ since,
as shown above, for this worldline\footnote{Notice however that $\nabla_{\delta x}a^{\alpha}\ne0$ and $\nabla_{\delta x}\Phid^{\alpha}\ne0$.}
$a^{\alpha}=0\Rightarrow\Phid^{\alpha}=0\Rightarrow P^{\alpha}=mU^{\alpha}$.
For such a motion, Eq.~\eqref{eq:DeltaF0} reads 
\begin{equation}
\nabla_{\delta x}F^{\alpha}=m\mathbb{E}_{\ \nu}^{\alpha}\delta x^{\nu}-\frac{1}{2}R_{\ \beta\mu\nu}^{\alpha}S^{\mu\nu}\CovU\delta x^{\beta}\quad,\label{eq:DeltaF}
\end{equation}
whose first term exactly cancels out the tidal term in (\ref{eq:Wdev12}),
i.e. one has then 
\begin{align}
\frac{{\rm D}^{2}\delta x^{\alpha}}{{\rm d}\tau^{2}}= & -\frac{1}{2m}R_{\ \beta\mu\nu}^{\alpha}S^{\mu\nu}\CovU\delta x^{\beta}-\nabla_{\delta x}\CovU\frac{\Phid^{\alpha}}{m}\ .\label{eq:Wdev2}
\end{align}
These expressions have the following interpretation. The first term
of the force variation~\eqref{eq:DeltaF} ensures that the worldlines
move at a constant distance, by counteracting the tidal force $-\mathbb{E}_{\ \nu}^{\alpha}\delta x^{\nu}$
in Eq.~\eqref{eq:Wdev12}, which ``tries'' to make the worldlines
diverge/converge. The second term of Eq.~\eqref{eq:DeltaF}, together
with the hidden momentum term $\nabla_{\delta x}\CovU(\Phid^{\alpha}/m)$
in Eq.~\eqref{eq:Wdev12}, which form Eq.~\eqref{eq:Wdev2}, are
responsible for the winding motion around the nonhelical centroid.

In a flat spacetime, such a winding---and hence the relative acceleration
between centroids---are solely due to the hidden momentum: ${\rm D}^{2}\delta x^{\alpha}/{\rm d}\tau^{2}=-\nabla_{\delta x}\CovU(\Phid^{\alpha}/m)$,
cf. Eq. \eqref{eq:Wdev2}. Curvature changes both $\Phid^{\alpha}$
and its derivative, but such a change is nevertheless compensated
by the first term of Eq.~\eqref{eq:Wdev2} {[}second term of Eq.~\eqref{eq:DeltaF}{]}
ensuring that, apart from an overall motion in the radial direction,
the trajectories are almost the same as in flat spacetime, as manifest
in Figs.~\ref{fig:RadialSpin},~\ref{fig:PolarSpin}.

In order to see how these things play out in the examples herein,
we first notice that the connecting vector $\delta x^{\alpha}$ is
simply related with the shift vector $\Delta x^{\alpha}$: \emph{for
worldlines infinitesimally close}, $\Delta x^{\alpha}=h_{\ \beta}^{\alpha}\delta x^{\beta}$;
that is, $\Delta x^{\alpha}$ is the projection of $\delta x^{\alpha}$
in the direction orthogonal to the basis worldline $z^{\alpha}(\tau)$
{[}see Eq. (\ref{eq:SpaceProj}){]}. Consider now the radial spin
case of Fig.~\ref{fig:RadialSpin}. The spin-curvature force exerted
on the helical centroids $\bar{z}^{\alpha}$ is, to leading post-Newtonian
order, $\vec{\bar{F}}\simeq-3M\vec{v}\times\vec{S}/r^{3}$ (e.g. Eq.
(56) of \cite{CostaNatario2014}, for a radial\footnote{For a helical centroid $\bar{z}^{\alpha}(\tau)$ the spin vector that
follows from \eqref{eq:STransferAprox} is actually not exactly radial,
since $\bar{z}^{\alpha}(\tau)$ is shifted from the radial geodesic;
that however amounts to corrections of order $\mathcal{O}(S^{2})$
in $\vec{F}$.} $\vec{S}$), pointing \emph{outwards}, in the direction of the shift
vector $\Delta\vec{x}$. Since the force $\vec{F}$ on the nonhelical
centroid $z^{\alpha}$ is zero, we have that $\vec{\bar{F}}=\vec{\bar{F}}-\vec{F}\equiv\Delta\vec{F}$;
and since $\Delta\vec{F}\simeq\nabla_{\Delta x}\vec{F}$, we see that
the force $\vec{\bar{F}}$ consists of two parts, which are the two
terms of \eqref{eq:DeltaF} (with $\delta\vec{x}\rightarrow\Delta\vec{x}$).
The explicit expression for the first term follows from approximating
$\Delta\vec{x}\simeq\vec{S}\times\vec{v}/m$, and using the expression
for $\mathbb{E}^{ij}$ in e.g. Eq. (88) of \cite{CostaNatarioZilhao},
leading to $m\mathbb{E}^{ij}\Delta x_{j}\simeq-M(\vec{v}\times\vec{S})^{i}/r^{3}=\vec{\bar{F}}/3$
(so it amounts to one third of the force). This term counteracts the
tidal force between $\bar{z}^{\alpha}$ and $z^{\alpha}$ (first term
of \eqref{eq:Wdev12}, with $\delta\vec{x}\rightarrow\Delta\vec{x}$),
which is \emph{compressive}, i.e., antiparallel to $\Delta\vec{x}$,
preventing the two worldlines from converging. To compute the second
term of Eq.~\eqref{eq:DeltaF}, we first note that $\CovU\delta x^{\beta}$
relates to the relative velocity $v^{\alpha}$ between \emph{infinitesimally
close} worldlines $\bar{z}^{\alpha}(\tau)$ and $z^{\alpha}(\tau)$
by (cf. Eq. (4.27) of \cite{EMMbook}) 
\[
v^{\alpha}=h_{\ \beta}^{\alpha}\nabla_{U}(h_{\ \gamma}^{\beta}\delta x^{\gamma})=h_{\ \gamma}^{\alpha}\nabla_{U}\delta x^{\gamma}+a^{\alpha}U_{\gamma}\delta x^{\gamma}\ ,
\]
reducing to $v^{\alpha}=h_{\ \gamma}^{\alpha}\nabla_{U}\delta x^{\gamma}$
in the present case that the basis worldline $z^{\alpha}(\tau)$ is
geodesic. To leading post-Newtonian order, one can thus write, for
the second term of \eqref{eq:DeltaF}, $-R_{\ jkl}^{\alpha}\epsilon_{\ \ m}^{kl}S^{m}v^{j}/2$,
yielding $-2M\vec{v}\times\vec{S}/r^{3}=2\vec{\bar{F}}/3$. Two aspects
of the latter term are worth mentioning: (i) unlike the first term
of the force variation \eqref{eq:DeltaF} (which is due to the dependence
of the force on the centroid's position), this one is due to the dependence
of the spin-curvature force on the centroid's velocity; (ii) it is
only for helical solutions of the MP SSC (where $\CovU\delta x^{\beta}\sim v\sim\mathcal{O}(\delta x/S)$)
that this term is non-negligible. Under other SSCs (see \cite{CostaNatario2014}),
$\CovU\delta x^{\beta}\sim\mathcal{O}(\Phid)\sim\mathcal{O}(S^{n})$,
with $n\ge1$, so $R_{\ \beta\mu\nu}^{\alpha}S^{\mu\nu}\CovU\delta x^{\beta}\lesssim\mathcal{O}(m\rho^{2})$
is negligible to dipole order. This means that, between centroids
of other spin conditions, the difference in the forces equals minus
the tidal term: $\nabla_{\delta x}F^{\alpha}=m\mathbb{E}_{\ \nu}^{\alpha}\delta x^{\nu}$,
and thus the relative acceleration between centroids is solely down
to the hidden momentum term (cf. Sec. 3.3 of \cite{CostaNatario2014}).

\subsection{Circular equatorial orbits in Schwarzschild spacetime}

A problem where the MP SSC is a convenient choice is that of circular
equatorial orbits (CEOs) of a spinning particle in stationary axisymmetric
spacetimes, where (as shall be discussed in more detail elsewhere
\cite{CircularMP}), it allows to obtain, in a very simple fashion,
the \emph{exact} \emph{analytical solutions} for CEOs. Here we briefly
present the procedure for the special case of the Schwarzschild spacetime.

One starts by taking the spin vector to be polar, $S^{\alpha}=S^{\theta}\partial_{\theta}^{\alpha}$.
The spin evolution equation (\ref{eq:MPFermi}) ensures, in this case,
that the components $S^{\alpha}$ remain constant as long as $U^{r}=U^{\theta}=0$
(as is the case for a CEO).

For a CEO, the four-velocity has the form 
\begin{equation}
U^{\mu}=U^{0}(\partial_{0}+\OmOrb\partial_{\phi});\qquad U^{0}=\left[1-\frac{2M}{r}-r^{2}\OmOrb^{2}\right]^{-1/2}\label{eq:Ansatz}
\end{equation}
where $\OmOrb\equiv U^{\phi}/U^{0}$ is the (constant) angular velocity.
We take this as an ansatz for the centroid 4-velocity, and shall now
show that it is compatible with the equation of motion for the centroid
under the MP SSC, 
\begin{equation}
ma^{\alpha}=F^{\alpha}-\frac{{\rm D}\Phid^{\alpha}}{{\rm d}\tau}\ ,
\end{equation}
with $F^{\alpha}$ and $\Phid^{\alpha}$ given by Eqs. \eqref{eq:TidaltensorF}
and \eqref{eq:hiddenMom}. The acceleration corresponding to (\ref{eq:Ansatz})
has only radial component, 
\begin{equation}
a^{r}=-\frac{(r-2M)\left[r^{3}\OmOrb^{2}-M\right](U^{0})^{2}}{r^{3}}\quad.\label{eq:EqMotionMP}
\end{equation}
Now, for such 4-velocity and acceleration, and a polar spin vector
$S^{\alpha}=S^{\theta}\partial_{\theta}^{\alpha}$, both the spin-curvature
force \eqref{eq:TidaltensorF} and the covariant derivative of the
hidden momentum along $U^{\alpha}$ have, as only nonvanishing components,
\begin{equation}
F^{r}=\frac{3M(r-2M)S^{\theta}\OmOrb(U^{0})^{2}}{r^{2}}\ ,\label{eq:F}
\end{equation}
\begin{equation}
\frac{{\rm D}\Phid^{r}}{{\rm d}\tau}=\frac{(r-2M)(3M-r)(M-r^{3}\OmOrb^{2})S^{\theta}\OmOrb(U^{0})^{4}}{r^{3}}\ .\label{eq:Dphid}
\end{equation}
They are purely radial, just like the acceleration; it then follows
from \eqref{eq:EqMotionMP} that finding CEOs reduces to solving for
$\OmOrb$ the radial equation 
\begin{equation}
ma^{r}+\frac{DP_{{\rm hid}}^{r}}{d\tau}-F^{r}=0\ .\label{eq:equation}
\end{equation}
This is a fourth order equation for $\OmOrb$, leading to four distinct
solutions. Their explicit (lengthy) expressions, obtained using \emph{Mathematica},
are given in \cite{Supplement}. Two of the solutions are spurious
and do not reduce to the circular geodesics for $S=0$. One of them
(or both, depending on $r$ and the parameters $M$ and $S$) is unphysical,
as its speed is supra-luminal. The other is a ``giant'' highly-relativistic
helical motion of radius $r$, whose speed approaches the speed of
light as $r\rightarrow\infty$. It does not correspond to an orbital
motion of the physical body: its velocity remains nonzero and highly
relativistic even for $M\rightarrow0$ (Minkowski spacetime), when
it becomes an helical solution of a giant body \emph{at rest} in the
static frame (i.e., $\vec{P}=0$ in such frame), similar to those
depicted in body 1 of Fig. \ref{fig:Determinacy}. More details on
these solutions shall be given in \cite{CircularMP}.

The remaining two solutions are the physically relevant ones, corresponding
to ``prograde'' and ``retrograde'' orbits (i.e., positive or negative
angular velocity $\dot{\phi}$, respectively), which reduce to circular
geodesics when $S=0$ (and whose velocity appropriately reduces to
zero when $M\rightarrow0$). All four solutions match numerical results
in \cite{Harmsetal2016}.

Finally, concerning the problem of CEO's under other spin conditions,
to our knowledge, exact, analytical solutions, have so far been obtained
only with the TD condition \cite{Hackman2014}, through lengthier
computations.\footnote{CEO's are obtained from the results in \cite{Hackman2014} through
the following algorithm: one expresses $P_{0}$ and $P_{\phi}$ in
terms of the conserved ``energy'' $E$ and ``angular momentum''
$J$ {[}Eqs. (37)-(38), (40)-(41) therein{]}; then use the $U-P$
relation (13), (22)-(24) to express $U^{\alpha}$ also terms of $E$
and $J$. The condition $U^{r}=0$ eventually leads to Eq. (47) on
the quantity (44), which is solved for $E$ and $J$. Substituting
back into Eqs. (40)-(41), (13), (22)-(24) therein, yields $P^{\alpha}$,
and, finally, the 4-velocity $U^{\alpha}$ of the circular orbits.}

\subsubsection{Helical centroids}

To study helical centroids $\bar{z}^{\alpha}(\bar{\tau})$ corresponding
initially (within the realm of the pole-dipole approximation) to the
same physical body as the one described by a given circular orbit
$z^{\alpha}(\tau)$, we shall consider radial shifts. This is achieved
by demanding in Eq.~\eqref{eq:vrel} the \emph{initial} relative
velocity of $\bar{z}^{\alpha}(\bar{\tau})$ with respect to $z^{\alpha}(\tau)$
to be azimuthal, $v^{\mu}=v^{t}\partial_{t}^{\mu}+v^{\phi}\partial_{\phi}^{\mu}$.
In practice one needs only to prescribe its magnitude $v\equiv\sqrt{v^{\alpha}v_{\alpha}}<1$
and sign; orthogonality to $U^{\alpha}$ then yields the explicit
components 
\begin{equation}
v^{\phi}=\pm\frac{v}{\sqrt{g_{\phi\phi}+\OmOrb^{2}(g_{\phi\phi})^{2}/g_{00}}},\qquad v^{0}=-\frac{\OmOrb g_{\phi\phi}}{g_{00}}v^{\phi}\ .\label{eq:vrelCirc}
\end{equation}
This leads to $\bar{U}^{\mu}|_{z}=\bar{U}^{0}|_{z}\partial_{0}^{\mu}+\bar{U}^{\phi}|_{z}\partial_{\phi}^{\mu}$
and to a radial shift vector, cf. Eq.~\eqref{eq:DeltaxCov}, 
\begin{equation}
\Delta x^{\alpha}|_{{\rm in}}=-\frac{S_{\ \beta}^{\alpha}\bar{U}^{\beta}|_{z}}{\bar{m}}=\gamma\frac{\epsilon_{\ \beta\gamma\delta}^{\alpha}S^{\gamma}U^{\delta}v^{\beta}}{\bar{m}}=\Delta x^{r}\delta_{r}^{\alpha}\ ,\label{eq:ShiftCircular}
\end{equation}
where $\gamma\equiv-U_{\beta}\bar{U}^{\beta}|_{z}$, cf. Eq.~\eqref{eq:vrel}.

Herein no approximation will be made, so the starting point $\bar{z}^{\mu}$
is obtained by exact application of the exponential map $\bar{z}^{\mu}=\exp_{z}^{\mu}(\Delta x)$.
For that, we first note that the radial lines $\theta={\rm const}.$,
$\phi={\rm const.}$, $t={\rm const.}$ are spatial geodesics in Schwarzschild's
spacetime.\footnote{This is seen from the geodesic equation which, for a radial tangent
vector $\eta^{\alpha}\equiv\delta_{r}^{\alpha}dr/ds$, reads $d\eta^{r}/ds+\Gamma_{rr}^{r}(\eta^{r})^{2}=0\Leftrightarrow d\eta^{r}/dr+\Gamma_{rr}^{r}\eta^{r}=0$.
This is a first order ODE with solution $\eta^{r}=C\sqrt{1-\frac{2M}{r}}$,
being $C$ an arbitrary constant ($C=1$ if $s$ is chosen as the
arclength of the curve, case in which $\eta^{\alpha}\eta_{\alpha}=1$).} Along such a geodesic, the line element is $dl^{2}=g_{rr}dr^{2}$.
Hence, being $z^{\alpha}$ and $\bar{z}^{\alpha}$ points along that
curve, by definition of the exponential map, the arclength of the
segment between $z^{\alpha}$ and $\bar{z}^{\alpha}$ equals the magnitude
of $\Delta x^{\alpha}$: $\|\Delta x^{\alpha}\|=\mp\int_{r}^{\bar{r}}\sqrt{g_{rr}}dr$,
the $\text{+}$ ($-$) sign applying when $\bar{r}>r$ ($\bar{r}<r$).
Integrating this leads to 
\begin{align}
 & \bar{r}\sqrt{1-\frac{2M}{\bar{r}}}+M\ln\left[\bar{r}-M+\bar{r}\sqrt{1-\frac{2M}{\bar{r}}}\right]=\nonumber \\
 & \pm\|\Delta x^{\alpha}\|+r\sqrt{1-\frac{2M}{r}}+M\ln\left[r-M+r\sqrt{1-\frac{2M}{r}}\right]\label{eq:Exactr}
\end{align}
which is an equation that yields $\bar{r}$ (thus $\bar{z}^{\alpha}$),
given the values of $\|\Delta x^{\alpha}\|$ and $r$, to be solved
numerically.

The parallel transport of the moments $P^{\alpha}$ and the transformed
spin tensor $\bar{S}^{\alpha\beta}|_{z}$, Eqs. (\ref{eq:TransferP})-(\ref{eq:TransferS}),
shall also be calculated exactly. Let $\eta^{\alpha}=\delta_{r}^{\alpha}dr/ds$
denote the tangent vector to the spatial geodesic $c^{\alpha}(s)$
connecting $z^{\alpha}$ to $\bar{z}^{\alpha}$. It is easy to check
that the parallel transport conditions $\nabla_{\eta}P^{\alpha}=0$
and $\nabla_{\eta}\bar{S}^{\alpha\beta}=0$ are satisfied if these
tensors have constant components in the orthonormal tetrad $\mathbf{e}_{\hat{\alpha}}$,
tied to the Schwarzschild basis vectors $\partial_{\alpha}$, defined
by 
\[
\mathbf{e}_{\hat{i}}=(1/\sqrt{g_{ii}})\partial_{i},\quad\mathbf{e}_{\hat{0}}=(1/\sqrt{-g_{00}})\partial_{0}\quad.
\]
That is, when one has 
\[
P^{0}=P^{\hat{0}}/\sqrt{-g_{00}},\quad P^{i}=P^{\hat{i}}/\sqrt{g_{ii}}\quad,
\]
\[
\bar{S}^{ij}=\bar{S}^{\hat{i}\hat{j}}/\sqrt{g_{ii}g_{jj}},\quad\bar{S}^{i0}=\bar{S}^{\hat{i}\hat{0}}/\sqrt{-g_{00}g_{ii}}\quad,
\]
with $P^{\hat{\alpha}}$ and $\bar{S}^{\hat{\alpha}\hat{\beta}}$
\emph{constant}. This leads to the relations 
\begin{align}
 & P^{i}|_{\bar{z}}=P^{i}|_{z}\frac{\sqrt{g_{ii}}|_{z}}{\sqrt{g_{ii}}|_{\bar{z}}},\quad P^{0}|_{\bar{z}}=P^{0}|_{z}\frac{\sqrt{-g_{00}}|_{z}}{\sqrt{-g_{00}}|_{\bar{z}}}\quad,\label{eq:ExactTransportP}\\
 & \bar{S}^{ij}|_{\bar{z}}=\bar{S}^{ij}|_{z}\frac{\sqrt{g_{ii}g_{jj}}|_{z}}{\sqrt{g_{ii}g_{jj}}|_{\bar{z}}},\quad\bar{S}^{i0}|_{\bar{z}}=\bar{S}^{i0}|_{z}\frac{\sqrt{-g_{ii}g_{00}}|_{z}}{\sqrt{-g_{ii}g_{00}}|_{\bar{z}}}\ \quad.\label{eq:ExactTransportS}
\end{align}
For given values of $z^{\alpha}$, $P^{\alpha}$, and $S^{\alpha\beta}$
of the basis centroid, and of the magnitude $v$ of the initial relative
velocity of $\bar{z}^{\alpha}$with respect to $z^{\alpha}$, Eqs.~\eqref{eq:vrelCirc}-\eqref{eq:ExactTransportS}
yield the initial data needed for helical solutions. The initial data
are then numerically evolved by the equations of motion (\ref{eq:EqsMotion})
{[}i.e., Eqs. \eqref{Papa-p}-\eqref{eq:Spinevol} plus the momentum-velocity
relation~\eqref{eq:momentumVelMP}{]}. 

\begin{figure}
\includegraphics[width=1\columnwidth]{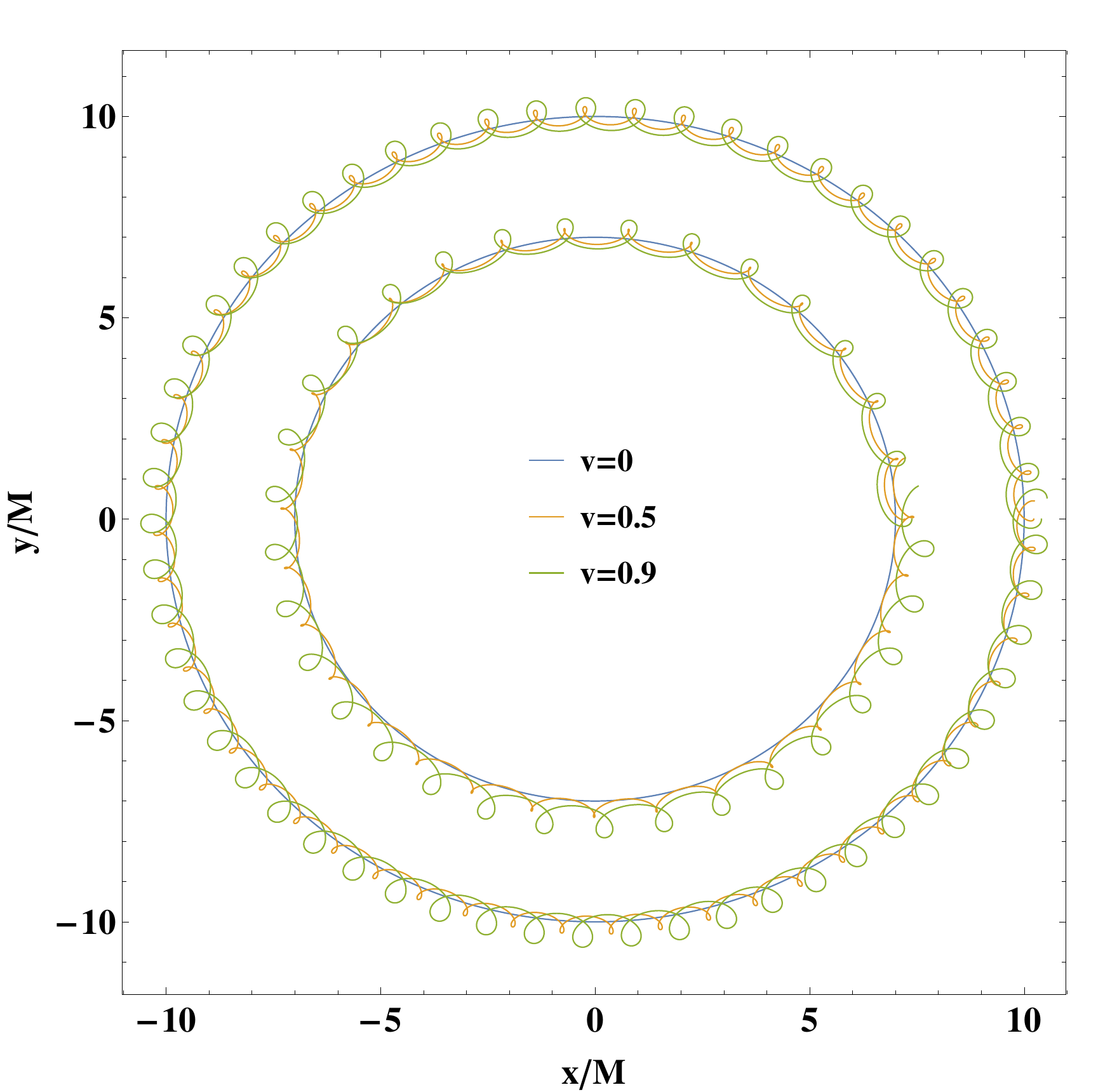}

\protect\caption{\label{fig:Circular_7_10}Circular ``prograde'' orbits for $r=7M$
and $r=10M$ (blue lines) and the helical solutions representing the
same physical motions, corresponding to two different values of the
relative velocity in Eq. (\ref{eq:vrelCirc}) ($v=0.5$ and $v=0.9$).
All the trajectories start at $\phi=0$, and only the first lap about
the black hole is plotted. The spin angular momentum of the body,
measured \emph{about the nonhelical centroid}, has magnitude $S=0.5mM$.
As the motion progresses, the helices start ``detaching'' one from
another; for $r=7M$ the trajectories visibly diverge outside the
body's minimal worldtube (of radius $R_{{\rm Moller}}=S_{\star}/\mathcal{M}\simeq S/m=0.5M$).
This signals a breakdown of the approximation scheme for these centroids.}
\end{figure}

\begin{figure}
{\includegraphics[height=0.9\columnwidth]{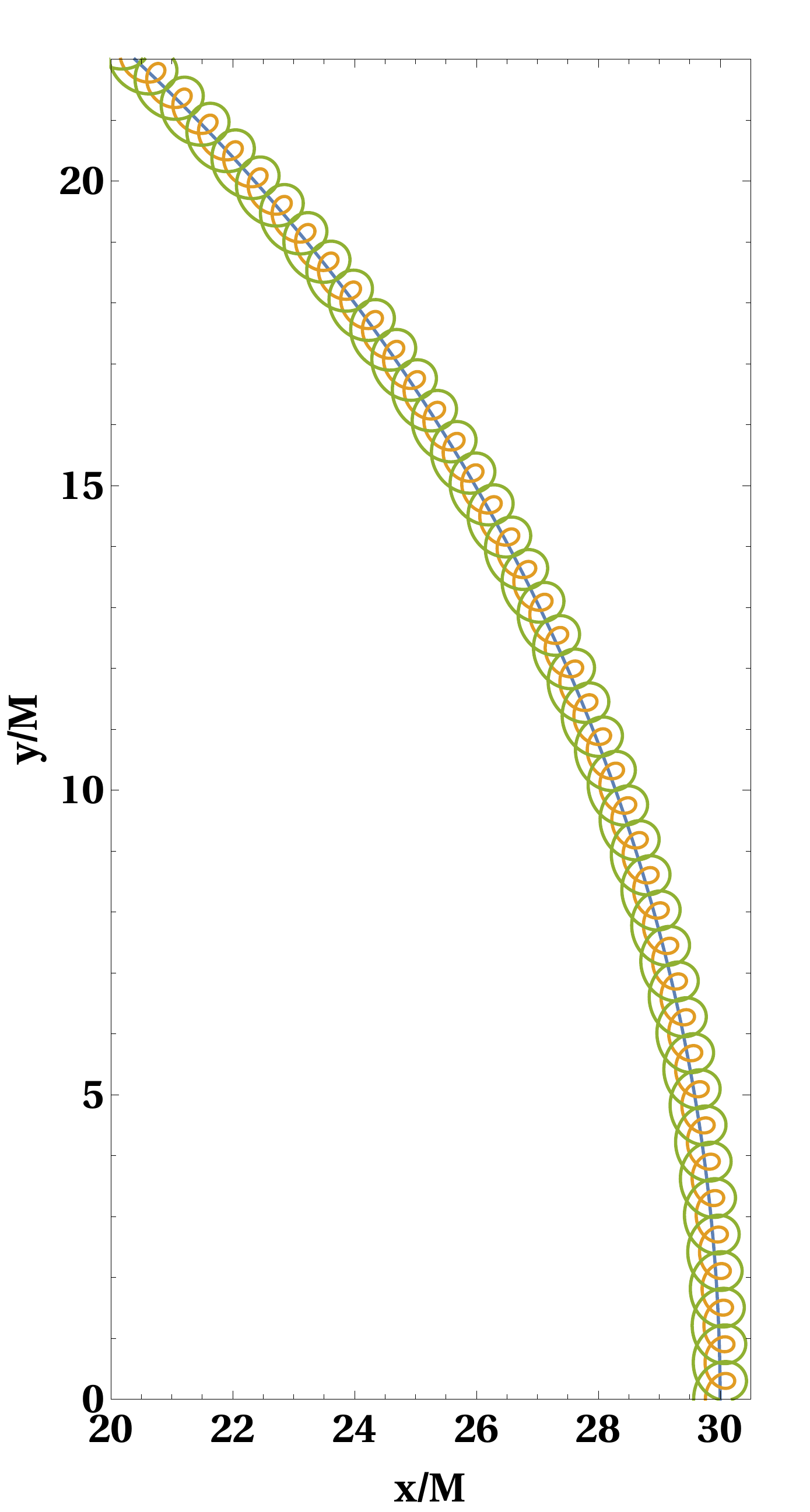}~\includegraphics[height=0.9\columnwidth]{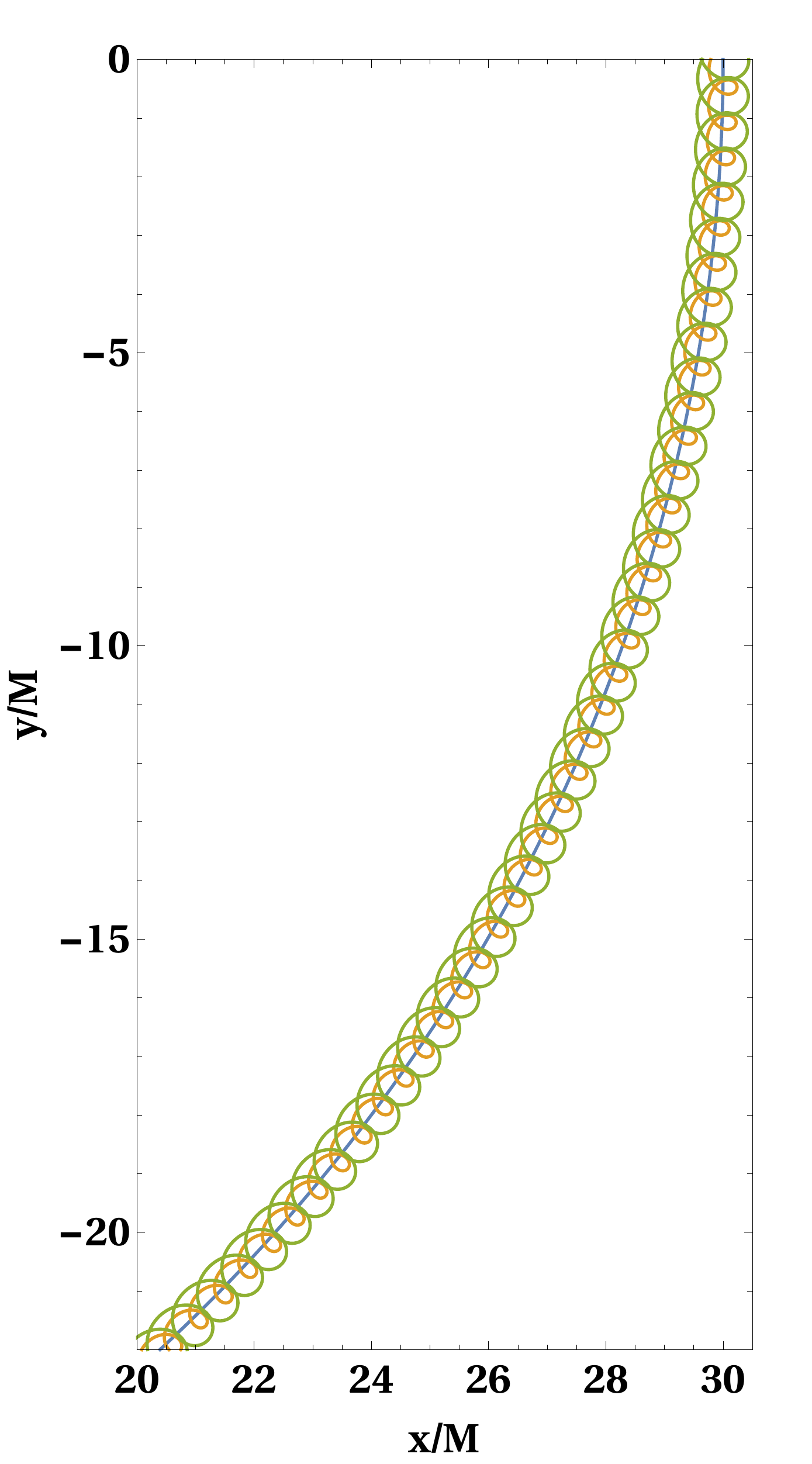}}

\protect\caption{\label{fig:Circular_30}Similar plots to Fig. \ref{fig:Circular_7_10},
now for $r=30M$. (Due to size constrains, only the initial and final
segments of the first lap are shown. For the full plot, see \cite{Supplement2}).}
\end{figure}

\begin{figure}
\includegraphics[width=0.5\columnwidth]{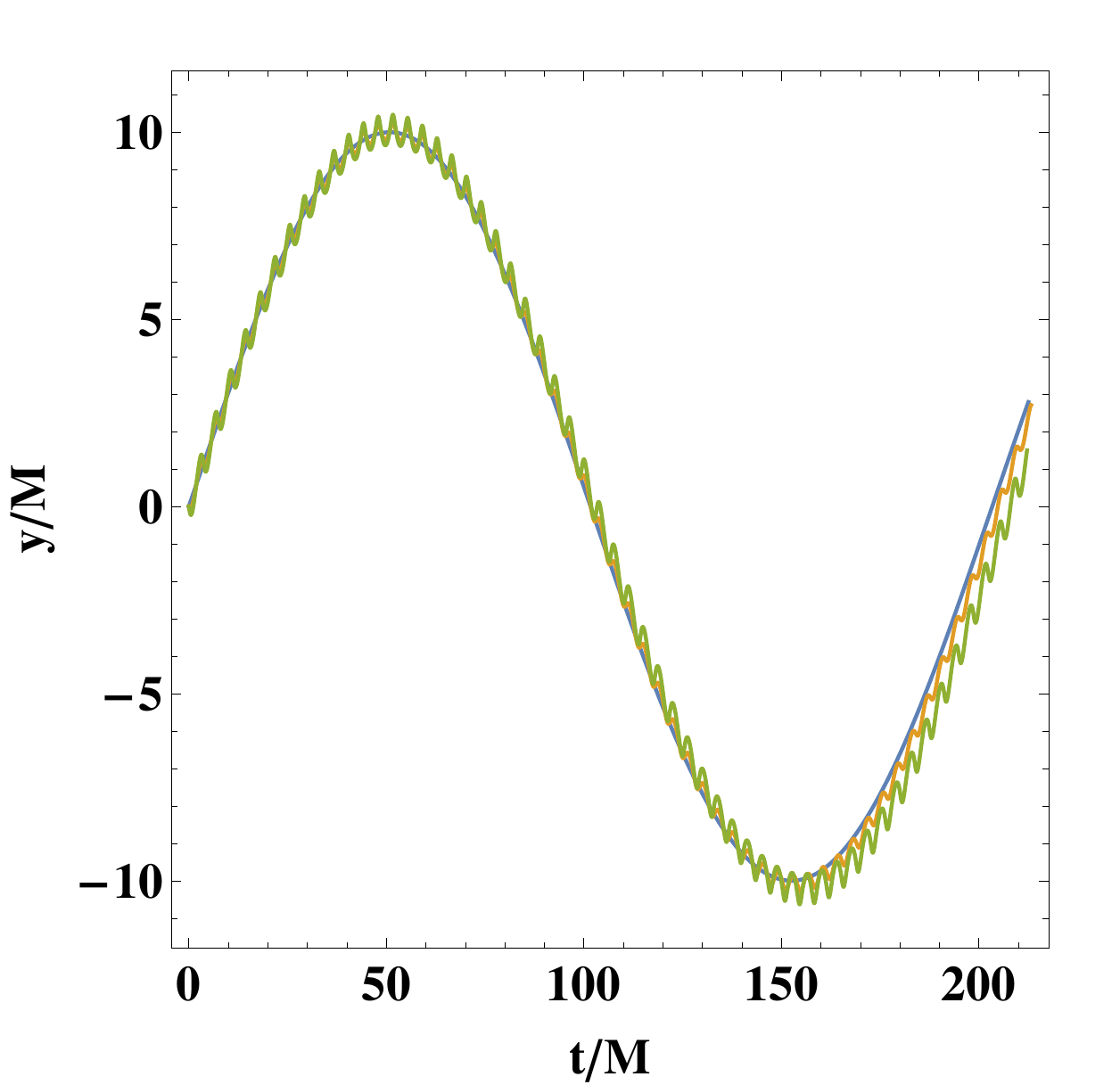}\includegraphics[width=0.5\columnwidth]{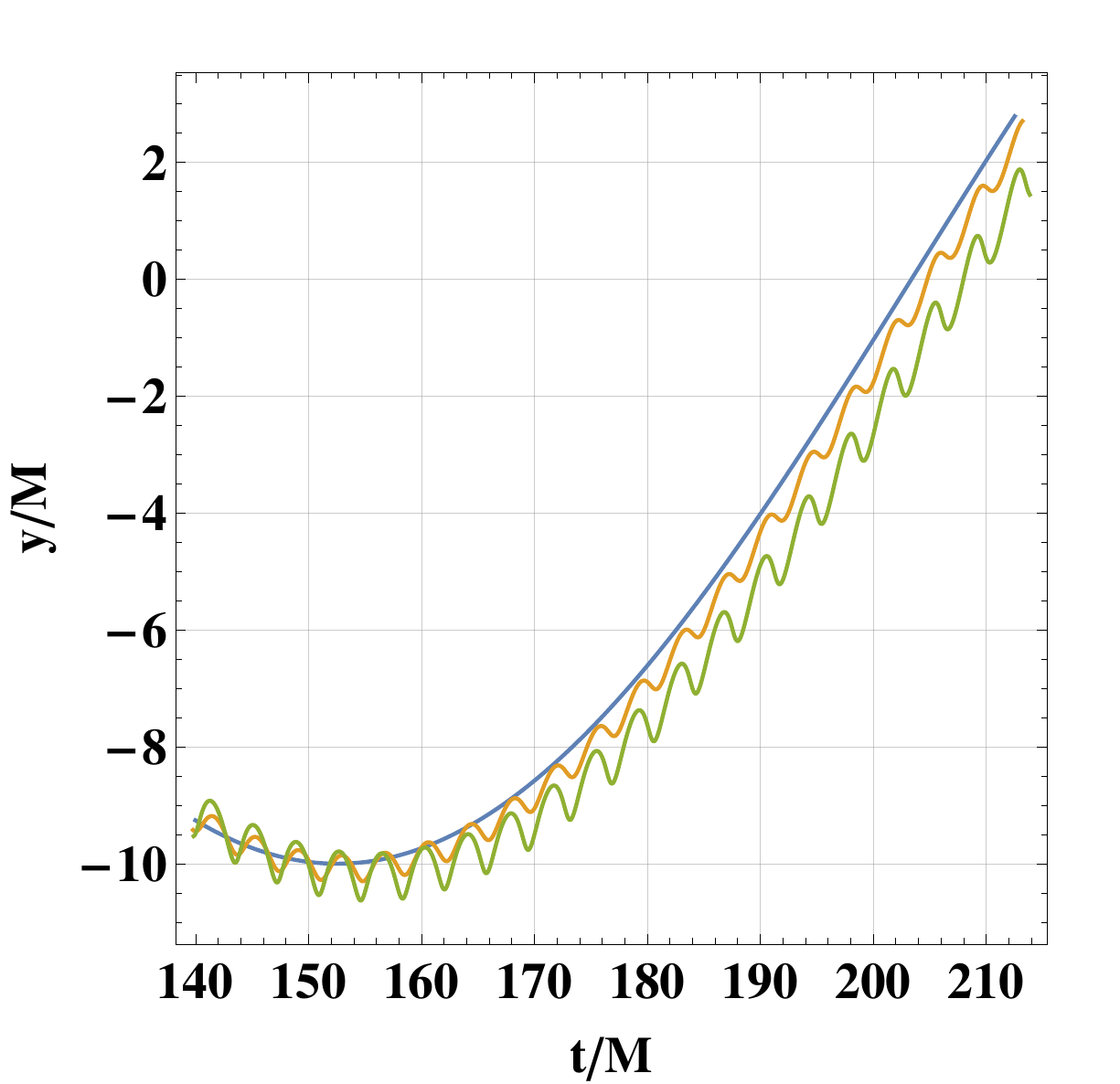}
\protect\caption{\label{fig:Spacetime_plot} The left panel shows how the $y/M$ evolves
during the first lap with respect to the coordinate time $t$ for
the trajectories in Fig.~\ref{fig:Circular_7_10} corresponding to
$r=10M$. The color scheme is the same as in Fig.~\ref{fig:Circular_7_10}.
The right panel is a closeup of the final part of the first lap. The
separation between points of the different worldlines for the same
instant $t$ becomes larger than the body's disk of centroids (of
radius $R_{{\rm Moller}}=S_{\star}/\mathcal{M}\simeq S/m=0.5M$),
signaling a breakdown of the approximation scheme for these centroids.}
\end{figure}

They are plotted, together with the corresponding nonhelical solutions,
in Figs.~\ref{fig:Circular_7_10}-\ref{fig:Circular_30}, for $r=7M$,
$r=10M$ and $r=30M$. Initially the helices are winding about, approximately,
the nonhelical worldline, much like in the way they do in flat spacetime
(see Sec. 1 of \cite{Supplement2}), as one would expect for different
worldlines of the same body (recall discussion in Sec.~\ref{sub:Radial-fall-in}).
However, as the motion progresses, the plots show that the helices
start detaching one from another, in the sense that the ``peaks''
do not meet. The effect is larger the closer the orbit is to the horizon
(i.e., the stronger the field). For $r=7M$ they visibly diverge outside
the spatial tube swept by the body's minimum size, which is the size
of its disk of centroids, of radius $R_{{\rm Moller}}=S_{\star}/\mathcal{M}\simeq S/m$,
cf. Eq. (\ref{eq:Rmoller}). Here $m=-P^{\alpha}U_{\alpha}$ and $S$
are, respectively, the mass and spin corresponding to the \emph{nonhelical}
centroid $z^{\alpha}$, and in the approximate equality $S_{\star}/\mathcal{M}\simeq S/m$
we noted that $m=\gamma(P,U)\mathcal{M}$ {[}where $\gamma(P,U)$
is the Lorentz factor between $U^{\alpha}$ and $P^{\alpha}/\mathcal{M}${]},
used the flat spacetime relation (\ref{eq:SSstarvector}) to estimate
$S\simeq S_{\star}/\gamma(P,U)$, and finally noted that $\gamma(P,U)\approx1$,
since $U^{\alpha}$ is very nearly parallel to $P^{\alpha}$ for a
nonhelical centroid (e.g. for $r=10M$, $\gamma(P,U)-1=10^{-8}$).
For $r=30M$ and $r=10M$, the effect is less pronounced, and the
trajectories of the helical centroids stay contained within a spatial
tube seemingly consistent with the size of the body's disk of centroids.
Although in Figs. \ref{fig:Circular_7_10}-\ref{fig:Circular_30}
only one lap is depicted, the situation does not change significantly
after several laps (see additional plots in \cite{Supplement2}).
Nevertheless, even in these cases, simultaneous points (in the sense
of having the same coordinate time $t$) on different worldlines become
separated, after some time, by ``illegal'' shifts, larger than the
body's Møller radius $R_{{\rm Moller}}$. This is shown by the spacetime
plot of position versus coordinate time $t$ in Fig.~\ref{fig:Spacetime_plot}.
The plot also reveals that the helical orbits have an overall orbital
velocity slightly smaller than the nonhelical centroid. The effect
grows with the radius \emph{of the helix}, and it is not affected
on whether the initial shift points inwards or outwards (cf. additional
plots in \cite{Supplement2}).

Now, the transition rules between centroids devised in Sec.~\ref{sec:Different-solutions-correspondin},
as discussed therein, require $\lambda=\|\mathbf{R}\|\rho^{2}\ll1$.
As mentioned above, in order to have a finite spin $S$, a body must
have a minimum radius $\rho\ge R_{{\rm Moller}}\simeq S/m$; estimating
the Riemann tensor magnitude by $\|\mathbf{R}\|\simeq M/r^{3}$, we
have 
\[
{\displaystyle \lambda\simeq\|\mathbf{R}\|R_{{\rm Moller}}^{2}\sim\frac{S^{2}}{m^{2}}\frac{M}{r^{3}}=\left(\frac{S}{mM}\right)^{2}\left(\frac{r}{M}\right)^{-3}\quad.}
\]
We are using $S=0.5mM$, so, for $r=30M$ this yields $\lambda\sim10^{-5}$,
and, for $r=10M$ and $r=7M$, $\lambda\sim10^{-4}$, which well satisfies
the restriction $\lambda\ll1$. The illegal shifts are then likely
down to a breakdown of the pole-dipole approximation itself --- more
precisely, \emph{of the assumption that one can represent the same
body through different centroids, while at the same time keeping a
(dipole order) cutoff in the multipole expansions}. This is an unavoidable,
basic feature, that arises already in Newtonian mechanics (or electromagnetism),
when one describes an extended body through different representative
points.

Let us recall the Newtonian problem which is enlightening for the
problem at hand. Consider a homogeneous spherical body in Newtonian
mechanics. It is exactly a monopole body only with respect to one
point (the center of mass $z^{i}$); with respect to any other point
$z'^{i}$, it will have dipole, quadrupole, and (infinite) higher
order moments. Under a nonuniform gravitational field $\vec{G}(x)$,
the monopole force $m\vec{G}(z')$ with respect to $z'^{i}=z^{i}+\Delta x^{i}$
is different from the one at $z^{i}$, $m\vec{G}(z)$. That difference
is, however, exactly compensated by the dipole, quadrupole, ...$n$-pole
forces that arise at $z'^{i}$, so that the total Newtonian force
is the same in both cases, see Sec.~3.3 of \cite{CostaNatario2014}
for more details. The larger part of the compensation comes from the
dipole force $\vec{F}_{{\rm dip}}=-m\Delta\vec{x}\cdot\nabla\vec{G}$,
and a smaller part from the higher order moments. When one truncates
the expansion at a finite order, the compensation is not perfect.
Then the forces on the two points will no longer be exactly the same,
and the trajectories obtained generically will end up diverging.

The relativistic problem herein is analogous, only now the two points
$z^{\alpha}$ and $z'^{\alpha}$ are both centers of mass, and instead
of the gradient of the monopole force $m\vec{G}$ (which has no place
in general relativity) we talk about tidal forces, cf. Sec.~\ref{sub:Radial-fall-in}.
Let us assume that the body is well approximated by a pole-dipole
particle \emph{with respect to the nonhelical centroid}; i.e, it is
nearly ``spherical'' \cite{CostaNatarioZilhao}, and centered at
$z^{\alpha}$. When we shift to the helical centroid $z'^{\alpha}$
via Eqs.~\eqref{eq:ShiftCircular}-\eqref{eq:Exactr} and \eqref{eq:ExactTransportP}-\eqref{eq:ExactTransportS},
only the momentum $P^{\alpha}$ (of monopole order) and $S^{\alpha\beta}$
(dipole order) are adjusted. Thus, we are neglecting the quadrupole
and higher order moments that such shift generates. For a free particle
in flat spacetime this has no consequence in the dynamics. In a curved
spacetime however the gravitational field couples to such moments,
and the corresponding forces are needed for a full consistency of
the solutions.

For the nonhelical centroid having $v=0.9$ in Figs.~\ref{fig:Circular_7_10}-\ref{fig:Circular_30}
(the one for which the shift from $z^{\alpha}$ is larger, $\Delta x=0.9R_{{\rm Moller}}$),
the quadrupole force is of the order 
\[
F_{{\rm Q}}\sim m\|R_{\alpha\beta\gamma\delta,\lambda}\|R_{{\rm Moller}}^{2}\sim mMR_{{\rm Moller}}^{2}/r^{4}
\]
(cf. e.g., Eq.~(43) of \cite{Madore:1969}, Eq.~(7.4) of \cite{Dixon1970I}).
The change in the spin-curvature force in shifting from the nonhelical
centroid to the helical centroid for $v=0.9$ is of the order 
\[
\Delta F\sim m\|\mathbb{E}_{\ \nu}^{\alpha}\|R_{{\rm Moller}}\sim mMR_{{\rm Moller}}/r^{3}\quad,
\]
cf. Eq. (\ref{eq:DeltaF}). Thus, 
\[
\frac{F_{{\rm Q}}}{\Delta F}\sim\frac{R_{{\rm Moller}}}{r}\quad\left(\sim\frac{S}{mr}\right)\ \quad.
\]
In most astrophysical systems $R_{{\rm Moller}}\ll r$, so the quadrupole-force
correction is negligible compared to the spin-curvature one ($\Delta F$),
and it is therefore appropriate to shift between worldlines ignoring
quadrupole and higher moments, through the method proposed herein.

In the examples of Figs.~\ref{fig:Circular_7_10}-\ref{fig:Circular_30},
we are considering a spin magnitude $S=0.5mM$, so $F_{{\rm Q}}/\Delta F\sim0.5M/r$.
For $r=30M$, we have $F_{{\rm Q}}/\Delta F\sim0.01$, and for $r=7M$,
$F_{{\rm Q}}/\Delta F\sim0.1$, i.e. the quadrupole force is only
one order of magnitude smaller than $\Delta F$ and the spin-curvature
force itself. Given these orders of magnitude, the neglect of the
quadrupole order correction $F_{{\rm Q}}$ is expected to be reflected
on the orbits, and is likely\footnote{The neglect of the quadrupole force in the pole-dipole approximation
seems also possibly the cause for the eventual divergence of the centroids
of different spin conditions for the same body outside its ``minimal
worldtube,'' that have been found in \cite{Semerak II}.} the cause for the detaching of the helices and the inconsistent separation
between centroids in Figs. \ref{fig:Circular_7_10} and \ref{fig:Spacetime_plot}.

Finally, we note that in the examples of radial fall in Sec.~\ref{sub:Radial-fall-in}
this effect also arose, but much less pronounced. Namely, there is
only a slight misalignment, close to the horizon, in the ``peaks''
of the helices in Figs.~\ref{fig:RadialSpin} (right bottom panel)
and \ref{fig:PolarSpin}. The likely reason is that these orbits are
too short-lived, especially in the stronger field region, for the
effect to manifest itself. (One can infer about the duration of the
motion, in comparison with the progress of the circular orbits, from
the number of helical loops, since the frequency of the helices is
roughly the same in both settings.)

\section{Conclusions}

This paper concerns the role of the spin supplementary condition in
the spinning-particle problem, focusing mainly on the Mathisson-Pirani
(MP) version of the condition, $S^{\alpha\beta}U_{\beta}=0$. We start
by showing that the MP SSC has an explicit, and very simple, momentum-velocity
relation. This result was long-sought in the literature, and once
even thought not to exist. We clarify the apparent paradox between
such definite relation and the fact that this SSC is degenerate, solving
the apparent conflicts in the literature. We also explain the differences
from other SSCs regarding the initial data required to uniquely specify
a solution. These differences are seen to stem from MP's peculiar
momentum-velocity relation, and a thorough physical interpretation
of this feature is provided. Then, we explicitly show how, for a given
body, this SSC yields infinitely many possible representative worldlines,
generalizing, for a curved spacetime, the flat spacetime analysis
made in \cite{Helical}. In the process we establish a method for
transition between different representative worldlines corresponding
to the same body in a curved spacetime.

To illustrate these features, we considered settings, in Schwarzschild
spacetime, where this SSC is a convenient choice. Namely, we consider
(i) the case of a body (whose bulk is) initially at rest, in which
case it makes immediately clear that the body moves radially, as its
nonhelical centroid follows a radial geodesic; and (ii) the case of
the circular equatorial orbits, where it yields a very simple way
of showing that such orbits are possible, and to obtain them \emph{analytically}.
We then compare the evolution of different centroids (helical and
nonhelical) given by the MP SSC. Such a comparison, for different
solutions corresponding to \emph{the same body}, is done here for
the first time. In the radial motions case, we have found that (apart
from an overall increase in radial velocity as the body approaches
the black hole, due to its gravitational field), the helices are very
similar to their flat spacetime counterparts, even though their description
is substantially different (e.g. due to the spin-curvature force).
This we physically interpreted using the worldline deviation equation
for the congruence formed by the worldlines of the centroids obeying
this SSC.

A centroid shift implies a change in the body's multipole moments;
but in a pole-dipole approximation only moments up to the dipole order
(i.e. $P^{\alpha}$ and $S^{\alpha\beta}$) are adjusted. In flat
spacetime this has no consequences. In a curved spacetime, however,
curvature couples to the higher order moments, so ignoring them leads
to the trajectories that can no longer be \emph{exactly} consistent.
Given this fact, the results (Figs.~\ref{fig:RadialSpin}-\ref{fig:PolarSpin})
show that the pole-dipole approximation holds surprisingly well in
the radial motion examples. On the other hand, the CEOs provide trajectories
lasting long enough, in a strong field region, to seemingly reveal
these limitations.

An important point to emphasize, regarding the helical motions, is
their nature as \emph{pure gauge} effects (in other words, ``noise'').
Contrary to some suggestions made in the literature, they are not
wrong or unphysical, but they do not contain any new physics either,
nor are they down to any mysterious forces: \emph{the physical body
they represent does not undergo any helical motion} (so no experiment
could ever detect it), which is but a spurious motion of the representative
worldlines that this SSC does not exclude. This is so in flat spacetime
as shown in \cite{Helical}; herein we show that the same principle
naturally holds in a curved spacetime, just requiring a more subtle
treatment. In particular, by using proper transition rules to ensure
that one is dealing with solutions corresponding to a given body,
the different solutions (helical or nonhelical) remain close and describe,
within the scope of the pole-dipole approximation, the same physics.

It is crucial to distinguish the physical, measurable effects (i.e.,
those that reflect in the actual motion of the body's bulk), from
the pure gauge ones: spin effects in general relativity are typically
small, frequently within the same order of magnitude as the superfluous
motions induced by some spin conditions. For instance, the pure gauge
centroid acceleration induced by the CP or NW SSC's is of the same
order of magnitude as that originating from the actual spin-curvature
force \cite{CostaNatario2014,BakerOConnel19741975}. In the case of
the helical solutions of the MP SSC, such as those exemplified in
Sec. \ref{sec:Examples}, it is even typically much larger \cite{AccelHelices}.

Concerning the practicality of the MP SSC, the situation is ambivalent.
In those special cases where it is easy, e.g., thanks to the symmetries
of the problem, to prescribe the nonhelical solution, such as the
cases in Sec.~ \ref{sec:Examples}, or the ones treated in \cite{CostaNatarioZilhao},
this SSC can be of advantage. It is also suitable for some approximate
treatments, namely linear in spin approximations, where setting the
nonhelical centroid amounts to simply additionally demanding the centroid's
4-velocity $U^{\alpha}$ to be parallel to the body's momentum $P^{\alpha}$.
This can be seen by noticing, from Eq. (\ref{eq:P_U_Dixon}), that,
\emph{for the centroid fixed by the Tulczyjew-Dixon SSC} ($S^{\alpha\beta}P_{\beta}=0$),
one has $P^{\alpha}=mU^{\alpha}+\mathcal{O}(S^{2})$, implying that,
to such accuracy, it satisfies as well the MP SSC, and therefore coincides
with a centroid of the latter (the nonhelical one, since the hidden
momentum, which is a necessary ingredient for the helical motions,
cf. Sec. \ref{sub:WorldlineDev}, vanishes in this case by definition).
By definition, it also coincides with a centroid of the OKS SSC (the
one set up by initially choosing $V^{\alpha}=P^{\alpha}/\mathcal{M}$).
One may actually argue that such an approximation is inherent to the
spirit of the pole-dipole approximation \cite{QuadPhid}. The same
method can be applied in post-Newtonian schemes. However, in the framework
of an ``exact'' approach, and in the generic case when it is not
clear how to set the initial conditions for a nonhelical motion, the
MP SSC should rather be avoided, because the helices are superfluous.
They are just an \emph{unnecessarily} complicated description of motions
that can be made simpler using other SSCs. Thus, future prospects
for a wider applicability of the MP SSC crucially relies on finding
a generic method for singling out its nonhelical solution \cite{Plyatsko}.

\section*{Acknowledgments}

We thank J. Natário for useful discussions. L.F.C. is funded by FCT/Portugal
through Grant No. SFRH/BDP/85664/2012. G.L-G is supported by Grant
No. GACR-17-06962Y of the Czech Science Foundation. O.S. was supported
from the Grant No. GACR-17-13525S of the Czech Science Foundation
which is acknowledged gratefully. This work was partially supported
by FCT/Portugal through the project UID/MAT/04459/2013.


\begin{thebibliography}{10}
\bibitem{Plyatsko}Attempts for accomplishing this purpose have been
proposed in \cite{Plyatsko_Nonhelical}, for the special cases of
Kerr and Schwarzschild spacetimes; however, no general method has
yet been found.
\end{thebibliography}
\end{document}